\documentclass[aps,pre,showpacs,amsfonts,amsmath, amssymb,twocolumn,floatfix, superscriptaddress]{revtex4-1}
\usepackage[final]{graphicx}
\usepackage{xcolor}
\usepackage{verbatim}
\usepackage{csquotes}
\usepackage{braket}
\usepackage{mathtools}
\usepackage{diagbox}
\usepackage[colorlinks=true, linkcolor=blue, citecolor=blue, urlcolor=blue]{hyperref}

\definecolor{mygreen}{HTML}{006400}
\newcommand{\vct}[1]{\mathbf{#1}}

\newcommand{\add}[1]{\textcolor{mygreen}{ #1}}

\newcommand{\be}{\begin{equation}}
\newcommand{\ee}{\end{equation}}
\allowdisplaybreaks

\begin{document}
% HEADER_begin ---------------------------------------------

\title{Brownian systems perturbed by mild shear: Comparing response relations}

\author{Kiryl Asheichyk}
\email[]{asheichyk@bsu.by}
\affiliation{Department of Theoretical Physics and Astrophysics, Belarusian State University, 5 Babruiskaya St., 220006 Minsk, Belarus}
\thanks{Current address}
\affiliation{4th Institute for Theoretical Physics, Universit\"at Stuttgart, Pfaffenwaldring 57, 70569 Stuttgart, Germany}
\affiliation{Max Planck Institute for Intelligent Systems, Heisenbergstrasse 3, 70569 Stuttgart, Germany}
\author{Matthias Fuchs}
\email[]{matthias.fuchs@uni-konstanz.de}
\affiliation{Fachbereich Physik, Universit\"at Konstanz, 78457 Konstanz, Germany}
\author{Matthias Kr\"uger}
\email[]{matthias.kruger@uni-goettingen.de}
\affiliation{Institute for Theoretical Physics, Georg-August-Universit\"at G\"ottingen, 37073 G\"ottingen, Germany}

\begin{abstract}
We present a comprehensive study of the linear response of interacting underdamped Brownian particles to simple shear flow. We collect six different routes for computing the  response, two of which are based on the symmetry of the considered system and observable with respect to the shear axes. We include the extension of the Green-Kubo relation to underdamped cases, which shows two unexpected additional terms. These six computational methods are applied to investigate the relaxation of the response towards the steady state for different observables, where interesting effects due to interactions and a finite particle mass are observed. Moreover, we compare the different response relations in terms of their statistical efficiency, identifying their relative demand on experimental measurement time or computational resources in computer simulations. Finally, several measures of breakdown of linear response theory for larger shear rates are discussed.
\end{abstract}


\bibliographystyle{plain}

\maketitle
% HEADER_end -----------------------------------------------

% MAIN_PART_begin ------------------------------------------

\section{Introduction}
\label{sec:Intro}
Brownian motion plays a key role in dynamics of microparticles and living microorganisms, and fundamental research of Brownian systems is of high importance for practical applications in physics, biology, and medicine.

A natural way to understand any physical system, including a Brownian one, is to study its response to external perturbations. A small perturbation, where the response is linear in perturbation forces, already allows to draw important conclusions~\cite{Altland2010}. As a better alternative to the direct measurement of the linear response, linear response theory can be used, which allows to predict the response via two-point correlation functions of the unperturbed system~\cite{Altland2010}. This direct correspondence between the response and unperturbed fluctuations is known as the fluctuation-dissipation-theorem (FDT)~\cite{Altland2010, Marconi2008}. Starting from the famous Sutherland-Einstein-Smoluchowski relation~\cite{Altland2010, Marconi2008, Sutherland1905, Einstein1905, Smoluchowski1906}, a big progress in linear response theory has been made, mostly regarding generalization of FDT to various systems and perturbations~\cite{Altland2010, Marconi2008, Callen1951, Weber1956, Green1954, Kubo1957, Kubo1966, Kubo1991, Hansen2009}. In recent years, the focus has been moving towards the response of nonequilibrium systems~\cite{Harada2005, Speck2006, Blickle2007, Chetrite2008, Baiesi2009_1, Baiesi2009_2, Baiesi2010, Prost2009, Kruger2009, Kruger2010, Seifert2010_1, Seifert2010_2, Seifert2012, Warren2012, Szamel2017, Caprini2018, Asheichyk2019_1, DalCengio2019, DalCengio2021, Asheichyk2020, Caprini2021_1, Caprini2021_2}.

Yet linear response theory for equilibrium Brownian systems is still far from being understood comprehensively. Although it is known that one can use different response formulas to compute a certain response for a particular scenario~\cite{Marconi2008, Speck2006, Seifert2010_1, Asheichyk2019_1, Asheichyk2020, Asheichyk2019_2}, the comparison of the formulas (e.g., in terms of their statistical efficiency) was only very recently performed~\cite{Asheichyk2019_1, Asheichyk2020, Asheichyk2019_2}. Such a comparison can help choosing the best formula for a particular case, in order to minimize computational resources or experimental measurement time. Another important aspect, which has not been studied in detail, concerns the role of system properties in response computations. For example, certain symmetries of the considered system allow to compute the response via alternative ways~\cite{Asheichyk2019_1, Asheichyk2020, Asheichyk2019_2}. One more fact is that underdamped Brownian particles have been investigated much less frequently than overdamped ones, because the latter model is more simple yet widely applicable in many practical cases. Even if the particle inertia is formally considered~\cite{Baiesi2010, Warren2012, Asheichyk2020, Asheichyk2019_2, Rzehak2003, Holzer2010, Lander2012, Kahlert2012}, typically,  little attention is paid to its effect on the system response. On the other hand, it was proven that inertial effects play an important role at short time scales~\cite{Lukic2005, Blum2006, Li2010, Huang2011, Kheifets2014, Pusey2011}. In that respect, the relaxation dynamics of a perturbed system towards the steady state is expected to be very sensitive to the particle mass.

In this work, we address the three aforementioned aspects in linear response theory of Brownian motion. As for the perturbation, we consider simple shear flow, an important paradigmatic force driving the system out of equilibrium~\cite{Kruger2009, Kruger2010, Seifert2010_1, Warren2012, Asheichyk2019_1, Asheichyk2020, Asheichyk2019_2, Rzehak2003, Holzer2010, Lander2012, Kahlert2012, Orihara2011, Derksen1990, Fuchs2015, Ortman2012, Ziehl2009, Fuchs2005, Amann2013, Uspal2015, tenHagen2011, Li2017, Sandoval2014, Sandoval2018}. Based on the approaches of Refs.~\cite{Asheichyk2019_1, Asheichyk2020, Asheichyk2019_2}, we demonstrate in detail how the response to shear can be equivalently replaced by the response to a potential, thanks to certain symmetries (Sec.~\ref{sec:Symmetries}). In Sec.~\ref{sec:Six_ways}, we collect six different routes for obtaining the linear response, including  a new Green-Kubo relation for underdamped Brownian particles. These six methods are demonstrated in a concrete example of underdamped interacting particles, where interesting relaxation effects due to the particle inertia are discussed and the four response relations are analyzed in detail in terms of their statistical efficiency (Sec.~\ref{sec:Investigation}). Finally, some aspects regarding the difference between the linear and nonlinear responses are investigated (Sec.~\ref{sec:Nonlinear}).

\section{System}
\label{sec:System}
\begin{figure}[!t]
\includegraphics[width=0.8\linewidth]{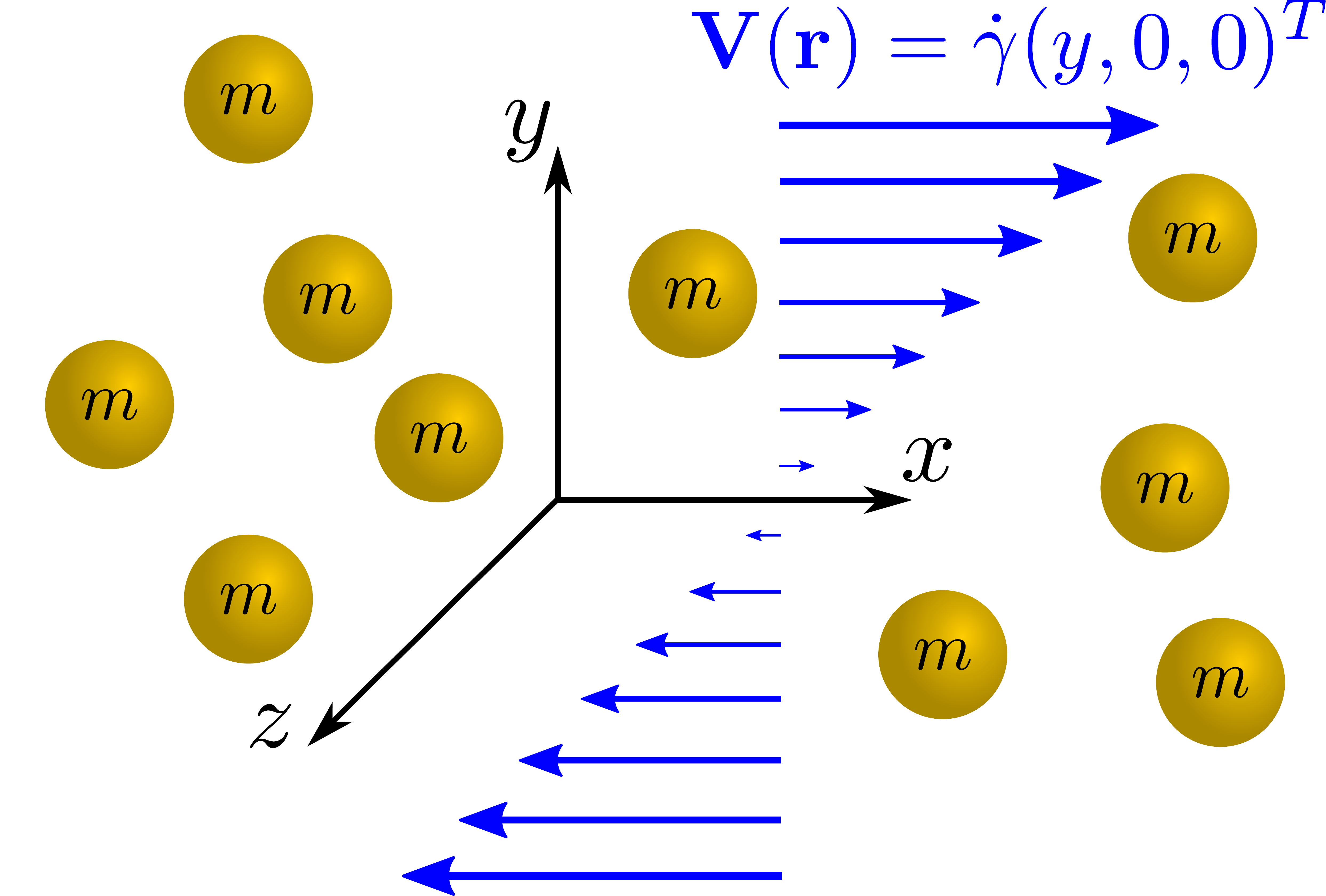}
	\caption{\label{fig:system}Schematic illustration of the considered system: Brownian particles (with a finite mass $ m $) perturbed by simple shear flow with the velocity profile $ \vct{V}(\vct{r}) = \dot{\gamma}(y, 0, 0)^T $ ($ \dot{\gamma} > 0 $ is the shear rate).}
\end{figure}

We consider a system of $ N $ interacting spherical Brownian particles, each with mass $ m $ and mobility $ \mu $, subject to external potential forces. Hydrodynamic effects are neglected. Notably, the chosen minimal model was found to predict the behavior of real Brownian systems both qualitatively and quantitatively~\cite{Lukic2005, Blum2006, Huang2011, Orihara2011, Henseler2010, Kreuter2012}. At time $ t = 0 $, the system is assumed to be in equilibrium. For time $ t > 0 $, it is perturbed by a simple shear flow with the velocity profile $ \vct{V}(\vct{r}) = \boldsymbol{\kappa}\cdot\vct{r} $, with the shear rate tensor $ \boldsymbol{\kappa} = \dot{\gamma}\hat{\vct{x}}\otimes \hat{\vct{y}}$ (where $ \dot{\gamma}$, $ \hat{\vct{x}} $, $ \hat{\vct{y}} $, and $ \otimes $ are the shear rate, the unit vectors, and the tensor product, respectively). Experimentally, such a flow can be realized in various ways, e.g., in a Couette device composed of two coaxial cylinders~\cite{Derksen1990, Fuchs2015}, in a sliding plate~\cite{Ortman2012} or cone-plate~\cite{Orihara2011} rheometer, or in a microfluidic device with two counter flows~\cite{Ziehl2009}. The dynamics of particle $ i $ is given by the Langevin equation for the velocity $ \vct{v}_i $ and the position $ \vct{r}_i $:
\begin{equation}
m\dot{\vct{v}}_i = \frac{1}{\mu}\boldsymbol{\kappa}\cdot\vct{r}_i -\frac{1}{\mu}\vct{v}_i + \vct{F}^{\textrm{int}}_i + \vct{F}^{\textrm{ext}}_i + \vct{f}_i, \ \ \ \ \dot{\vct{r}}_i = \vct{v}_i,
\label{eq:LE}
\end{equation}
where $ \vct{f}_i $ is a Gaussian white noise,
\begin{equation}
\langle \vct{f}_i(t) \rangle = 0, \ \ \ \ \langle \vct{f}_i(t) \otimes \vct{f}_j(t') \rangle = \frac{2k_{\textrm{B}}T}{\mu}\mathbb{I}\delta_{ij}\delta(t-t'),
\label{eq:RF}
\end{equation}
with $ k_{\textrm{B}} $, $ T $, and $ \mathbb{I} $ being the Boltzmann's constant, the temperature, and the identity matrix, respectively. $ \langle \cdots \rangle $ indicates average over the independent noise realizations. Interaction and external forces arise from the corresponding potentials: $ \vct{F}^{\textrm{int}}_i = -\boldsymbol{\nabla}_iU^{\textrm{int}} $ for internal (interaction) forces (where $ U^{\textrm{int}} $ is assumed to be radial),  $ \vct{F}^{\textrm{ext}}_i = -\boldsymbol{\nabla}_iU^{\textrm{ext}} $ for external forces. A schematic illustration of the system is given in Fig.~\ref{fig:system}.

\section{The role of symmetries: Response to shear via the response to a potential}
\label{sec:Symmetries}
The response of an observable $ A(t) $ is defined as $ \Delta A^{(\dot{\gamma})} \equiv {\left\langle A(t) \right\rangle}^{(\dot{\gamma})} - \left\langle A \right\rangle $ ($ t \geq 0 $), where the first and the second term on the right-hand side are the averages in the sheared and equilibrium system, respectively. Since we are interested in the \textit{linear} response, we consider only the term in $ \Delta A^{(\dot{\gamma})} $ which is linear in $ \dot{\gamma} $. In this section, we review the approach presented in Ref.~\cite{Asheichyk2019_2} to show that the linear response to shear is equivalent to the linear response to a potential if certain symmetry conditions are satisfied. The results of Ref.~\cite{Asheichyk2019_2} are complemented by new insights regarding these conditions: We illustratively demonstrate them based on the effect of system rotation and on the symmetry of equilibrium correlation functions.

\subsection{From the response to shear to the response to a potential}
\label{subsec:shear_pot}
Alongside with the shear perturbation force
\begin{equation}
\vct{F}_i^{\textrm{ptb}} = \frac{1}{\mu}\boldsymbol{\kappa}\cdot\vct{r}_i = \frac{\dot{\gamma}}{\mu}\left(y_i, 0, 0\right)^T,
\label{eq:F}
\end{equation}
consider a partner force~\cite{Asheichyk2020, Asheichyk2019_2, Dhont1996}
\begin{equation}
\vct{G}_i^{\textrm{ptb}} = \frac{1}{2\mu}\left(\boldsymbol{\kappa}^T - \boldsymbol{\kappa}\right)\cdot\vct{r}_i = \frac{\dot{\gamma}}{2\mu}\left(-y_i, x_i, 0\right)^T,
\label{eq:G}
\end{equation}
representing the half of the difference between shear flows in the $ y $ and $ x $ directions. Although both $ \vct{F}_i^{\textrm{ptb}} $ and $ \vct{G}_i^{\textrm{ptb}} $ are nonconservative forces, adding them together leads to a potential, $ \vct{F}_i^{\textrm{ptb}} + \vct{G}_i^{\textrm{ptb}} =  -\boldsymbol{\nabla}_iU^{\textrm{ptb}} $, where~\cite{Asheichyk2020, Asheichyk2019_2}
\begin{equation}
U^{\textrm{ptb}} = -\frac{\dot{\gamma}}{2\mu}\sum_{i=1}^Nx_iy_i.
\label{eq:Ushear}
\end{equation}
According to the superposition principle of the linear response, we get 
\begin{equation}
\Delta A^{(\dot{\gamma})} + \Delta A\big|_{\vct{G}^{\textrm{ptb}}} = \Delta A\big|_{U^{\textrm{ptb}}},
\label{eq:SPP}
\end{equation}
where $ \Delta A^{(\dot{\gamma})} = \Delta A\big|_{\vct{F}^{\textrm{ptb}}} $, and $  \Delta A\big|_{U^{\textrm{ptb}}} $ is the linear response to $ U^{\textrm{ptb}} $ in Eq.~\eqref{eq:Ushear}.

The main idea of Ref.~\cite{Asheichyk2019_2} is the following: If certain symmetry conditions are satisfied, then $ \Delta A\big|_{\vct{G}^{\textrm{ptb}}} = 0 $, and hence, 
\begin{equation}
\Delta A^{(\dot{\gamma})} = \Delta A\big|_{U^{\textrm{ptb}}},
\label{eq:ShearIsPotential}
\end{equation}
i.e., the response to shear is equivalent to the response to the potential in Eq.~\eqref{eq:Ushear}. This leads to the relation given in Eq.~\eqref{eq:LR_RK} below.

We note that the same idea can also be applied if the unperturbed system is out of equilibrium, as shown in Ref.~\cite{Asheichyk2019_1} for active Brownian particles.

\subsection{Symmetries}
\label{subsec:symmetries}
What are those symmetry conditions? This question can be answered using either of the following two facts. The first fact is that $ \vct{G}_i^{\textrm{ptb}} $ is orthogonal to the particle radius vector, 
\begin{equation}
\vct{G}_i^{\textrm{ptb}} \cdot \vct{r}_i = 0,
\label{eq:Gr}
\end{equation}
which means that $ \vct{G}_i^{\textrm{ptb}} $ is a rotation force. Because of its construction, it lies in the plane perpendicular to the $ z $ axis~\cite{Dhont1996}. As a result, in the linear order, the overall effect of $ \vct{G}_i^{\textrm{ptb}} $ is rotation of the system around the $ z $ axis.

The second fact is that the linear response to $ \vct{G}_i^{\textrm{ptb}} $ is given by an equilibrium correlation function of an observable $ A $ with a quantity which is antisymmetric under the interchange $ x_i \leftrightarrow y_i, \ v_{ix} \leftrightarrow v_{iy} $ (this notation implies the interchange for all particles)~\cite{Asheichyk2019_2, Risken1996, Kurchan1998}:
\begin{align}
\notag & \Delta A\big|_{\vct{G}^{\textrm{ptb}}} = \frac{1}{k_{\textrm{B}}T}\int_0^tdt'\left\langle A(t)\sum_{i=1}^N\vct{G}_i^{\textrm{ptb}}(t') \cdot \vct{v}_i(t')\right\rangle\\ 
& = \frac{\dot{\gamma}}{2k_{\textrm{B}}T\mu}\int_0^tdt'\left\langle A(t)\sum_{i=1}^N\left[x_i(t')v_{iy}(t')-y_i(t')v_{ix}(t')\right]\right\rangle.
\label{eq:G_LR}
\end{align}
Derivation of this formula is given in Appendix~\ref{app:KKE}. Note that the first line in Eq.~\eqref{eq:G_LR} tells that the linear response is determined by how much $ A $ is coupled to the work $ \int_0^tdt'\sum_{i=1}^N\vct{G}_i^{\textrm{ptb}}(t') \cdot \vct{v}_i(t') $ done by $ \vct{G}_i^{\textrm{ptb}} $ on the system~\cite{Asheichyk2020, Asheichyk2019_2}. 

Let us now introduce $ xy $ symmetry. Generally, we call any function of the phase space $ xy $ symmetric if this function remains unchanged under the interchange $ x_i \leftrightarrow y_i, \ v_{ix} \leftrightarrow v_{iy} $. A system is called $ xy $ symmetric if its distribution function is $ xy $ symmetric. For an equilibrium system, this is equivalent to saying that $ U^{\textrm{int}} $ and $ U^{\textrm{ext}} $ are $ xy $ symmetric.

Consider that the unperturbed system is $ xy $ symmetric. An example is Brownian spheres subject to gravity (acting along the $ z $ axis) and confined in a cubic box whose center lies on the $ z $ axis. Another example is a spherically symmetric system, i.e., $ U^{\textrm{int}} $ depends only on interparticle distances and $ U^{\textrm{ext}} $ is a function of $ |\vct{r}_i| $. All examples studied explicitly in this paper are spherically symmetric.

It is clear from physical grounds that a slow rotation of an $ xy $-symmetric system around the $ z $ axis does not change an $ xy $-symmetric observable if this observable depends only on particle positions. This argument is not so evident if $ A $ contains velocities, but it still holds thanks to the symmetry properties of equilibrium correlation functions. An equilibrium correlation function computed for an $ xy $-symmetric system is also $ xy $ symmetric (see Appendix~\ref{app:symm_eqcorr}). This means that 
\begin{equation}
\left\langle A(t)x_i(t')v_{iy}(t')\right\rangle = \left\langle \widetilde{A}(t)y_i(t')v_{ix}(t')\right\rangle,
\label{eq:symmetry_correlator}
\end{equation}
where $ \widetilde{A} $ is obtained from $ A $ by the interchange $ x_i \leftrightarrow y_i, \ v_{ix} \leftrightarrow v_{iy} $. If $ A $ is $ xy $ symmetric, then $ \widetilde{A} = A $, and hence $ \Delta A\big|_{\vct{G}^{\textrm{ptb}}} = 0 $ according to Eqs.~\eqref{eq:G_LR} and~\eqref{eq:symmetry_correlator}.

The above discussions allow us to repeat: For $ xy $-symmetric systems and observables, $ \Delta A\big|_{\vct{G}^{\textrm{ptb}}} = 0 $, and Eqs.~\eqref{eq:ShearIsPotential} and \eqref{eq:LR_RK} follow. This statement generalizes a known consideration that an antisymmetric part of the strain tensor (from which $ \vct{G}_i^{\textrm{ptb}} $ stems) drops out in the linear response for isotropic systems and spatial observables~\cite{Nagele1998}.
%\begin{equation}
%\Delta A^{(\dot{\gamma})} = \Delta A\big|_{U^{\textrm{ptb}}}.
%\label{eq:ShearIsPotential}
%\end{equation}

We note that, if the system is $ xy $ symmetric, many of the observables which have a finite linear response are also $ xy $ symmetric: for example, such observables are $ \sum_{i=1}^Nx_iy_i $, $ \sum_{i=1}^Nv_{ix}v_{iy} $, or the $ xy $ component of the stress tensor defined in Eq.~\eqref{eq:sigma}. There are however observables, which show a linear response, but for which Eq.~\eqref{eq:ShearIsPotential} fails. These couple positions and velocities, e.g., $ \sum_{i=1}^Nx_iv_{iy} $ (see Fig.~\ref{fig:Transient}).

\section{Collection of (six) ways to compute the linear response}
\label{sec:Six_ways}
In this section, we show that the linear response of Brownian particles to shear flow can be computed in at least six different ways: perturbing the system directly by shear or alternatively by the shear potential, or using one of the four response relations [Eqs.~\eqref{eq:LR_TA},~\eqref{eq:LR_RF},~\eqref{eq:LR_GKU}, and~\eqref{eq:LR_RK}]. Notably, Eq.~\eqref{eq:LR_GKU} has not been given in the literature.

\subsection{Perturbing the system}
\label{subsec:Perturbing}
The direct way to compute $ \Delta A^{(\dot{\gamma})} $ is to apply shear. Alternatively, according to Eq.~\eqref{eq:ShearIsPotential}, one can perturb the system by the shear potential~\eqref{eq:Ushear} if the system and $ A $ are $ xy $ symmetric.

\subsection{Computing equilibrium correlation functions: Linear response formulas}
\label{subsec:LR}
Linear response theory gives the response in terms of equilibrium correlation functions. Perturbing the system by a time independent potential $U^{\textrm{ptb}}$ switched on at time $t=0$, one has, what we call in the following, the equilibrium FDT~\cite{Baiesi2009_2, Baiesi2010, Risken1996}
\begin{equation}
\Delta A = - \frac{1}{k_{\textrm{B}}T}\left[\langle AU^{\textrm{ptb}}\rangle - \langle A(t) U^{\textrm{ptb}}(0) \rangle\right].
\label{eq:FDT_pp}
\end{equation}
Formula~\eqref{eq:FDT_pp} follows immediately from Eq.~\eqref{eq:G_LR}, when replacing $ \vct{G}_i^{\textrm{ptb}} $ with $ -\boldsymbol{\nabla}_iU^{\textrm{ptb}} $. Since the shear perturbation is a nonconservative one (i.e., the force in Eq.~\eqref{eq:F} does not arise from a potential), the corresponding response relation cannot, in general, be given by the equilibrium FDT~\cite{Marconi2008, Kubo1966, Hansen2009, Asheichyk2019_2}.
%\begin{equation}
%\Delta A = - \frac{1}{k_{\textrm{B}}T}\left[\langle AU^{\textrm{ptb}}\rangle - \langle A(t) U^{\textrm{ptb}}(0) \rangle\right],
%\label{eq:FDT_pp}
%\end{equation}
%valid if an equilibrium system is perturbed by a potential $ U^{\textrm{ptb}} $.
Therefore, various forms of that relation are used in the literature.

One of those forms was already introduced in Eq.~\eqref{eq:G_LR}, stating that the linear response is related to the work done by the perturbation force on the system\add{~\cite{Asheichyk2019_2, Risken1996, Kurchan1998}}. In case of shear, we get [by replacing $ \vct{G}_i^{\textrm{ptb}} $ in Eq.~\eqref{eq:G_LR} with  $ \vct{F}_i^{\textrm{ptb}} $ given in Eq.~\eqref{eq:F}, see also Appendix~\ref{app:KKE}]
\begin{equation}
\Delta A^{(\dot{\gamma})} = \frac{\dot{\gamma}}{k_{\textrm{B}}T\mu}\int_0^tdt'\left\langle A(t)\sum_{i=1}^Ny_i(t')v_{ix}(t')\right\rangle.
\label{eq:LR_TA}
\end{equation}
Another response relation involves the random force $ f_{ix} $,
\begin{equation}
\Delta A^{(\dot{\gamma})} = \frac{\dot{\gamma}}{2k_{\textrm{B}}T}\int_0^tdt'\left\langle A(t)\sum_{i=1}^Ny_i(t')f_{ix}(t')\right\rangle.
\label{eq:LR_RF}
\end{equation}
Equation~\eqref{eq:LR_RF} can be derived using a path integral representation of the dynamics~\cite{Speck2006, Asheichyk2020}. It also follows from the response formula in terms of the Malliavin weight~\cite{Warren2012}.  Notably, the stochastic integrals in Eqs.~\eqref{eq:LR_TA} and~\eqref{eq:LR_RF} are independent of the convention for the stochastic calculus (i.e., Ito versus Stratonovich)~\cite{Cugliandolo2017, Gardiner2010, Wynants2010, Asheichyk2020}, because noise for different spatial directions is uncoupled.

The linear response can also be computed using the underdamped Green-Kubo relation (see derivations in Appendices~\ref{app:KKE} and~\ref{app:GKU_derivation})~\cite{Asheichyk2020},
\begin{align}
\notag & \Delta A^{(\dot{\gamma})} = \frac{\dot{\gamma}}{k_{\textrm{B}}T} \int_0^tdt'\left\langle A(t')\sigma_{xy}(0)\right\rangle\\
& + \frac{\dot{\gamma}m}{k_{\textrm{B}}T}\left[\left\langle A\sum_{i=1}^Ny_iv_{ix}\right\rangle - \left\langle A(t)\sum_{i=1}^Ny_i(0)v_{ix}(0)\right\rangle\right], 
\label{eq:LR_GKU}
\end{align}
where \cite{Hansen2009}
\begin{equation}
\sigma_{xy} = -\sum_{i=1}^N\left[mv_{ix}v_{iy} +\left(F_{ix}^{\textrm{int}}+F_{ix}^{\textrm{ext}}\right)y_i\right]
\label{eq:sigma}
\end{equation}
is the $ xy $ component of the stress tensor (generalized to include external forces). Note that, in case of central forces, it is $ xy $ symmetric. To our knowledge, formula~\eqref{eq:LR_GKU} has not been previously derived, and only special cases were known~\cite{Hess1983}. A formula of the same structure but for a potential perturbation has been derived in Ref.~\cite{Baiesi2010}. The last two terms in Eq.~\eqref{eq:LR_GKU} (i.e., the terms without integral) are unexpected, but indeed turn out to be required in the numerical examples provided below. They are absent when considering the overdamped limit given by the famous Green-Kubo relation~\cite{Fuchs2005, Dhont1996, Heyes1994, Lin2017}:
\begin{equation}
\Delta A^{(\dot{\gamma})} = \frac{\dot{\gamma}}{k_{\textrm{B}}T} \int_0^tdt'\left\langle A(t')\sigma_{xy}(0)\right\rangle,
\label{eq:LR_GK}
\end{equation}
where $ \sigma_{xy} $ is given by Eq.~\eqref{eq:sigma} without the term containing velocities. 

For formulas~\eqref{eq:LR_TA},~\eqref{eq:LR_RF}, and~\eqref{eq:LR_GKU}, we note that each of them directly follows from the two others, because the quantities they involve are related by the equation of motion~\eqref{eq:LE}. 

Finally, if the system and $ A $ are $ xy $ symmetric, Eq.~\eqref{eq:ShearIsPotential} holds, and the response is given by the equilibrium FDT~\cite{Asheichyk2019_2}
\begin{equation}
\Delta A^{(\dot{\gamma})} = \frac{\dot\gamma}{2k_{\textrm{B}}T\mu}\left[\left\langle A\sum_{i=1}^Nx_iy_i \right\rangle - \left\langle A(t)\sum_{i=1}^Nx_i(0)y_i(0) \right\rangle\right],
\label{eq:LR_RK}
\end{equation}
obtained by inserting the shear potential~\eqref{eq:Ushear} into Eq.~\eqref{eq:FDT_pp}, and also derived in Appendix~\ref{app:KKE}. We refer to formula~\eqref{eq:LR_RK} as sFDT (shear FDT or FDT for shear).

Response relations~\eqref{eq:LR_TA} and~\eqref{eq:LR_RF} are general, in a sense that they are the same for the underdamped and overdamped cases [in contrast to Eqs.~\eqref{eq:LR_GKU} and~\eqref{eq:LR_GK}] and that they have no symmetry restrictions [in contrast to Eq.~\eqref{eq:LR_RK}]. Compared to Eqs.~\eqref{eq:LR_GKU},~\eqref{eq:LR_GK}, and~\eqref{eq:LR_RK}, they however require the instantaneous stochastic velocity $ v_{ix} $ [Eq.~\eqref{eq:LR_TA}] or the stochastic force $ f_{ix} $ [Eq.~\eqref{eq:LR_RF})] as an input. These quantities are typically hard to measure in experiments of Brownian systems. %Equations~\eqref{eq:LR_TA} and~\eqref{eq:LR_RF} are thus hardly applicable in experiments. This might be one of the reasons why they appear less frequently in the literature than the Green-Kubo relation~\eqref{eq:LR_GK}. 
For $ xy $ symmetric systems and observables, the response relation~\eqref{eq:LR_RK} is advantageous %over the other formulas, 
because it contains no time integral and requires measurement of simple quantities. Further comparison of the response relations is given in Sec~\ref{sec:Investigation}, where they are demonstrated in a concrete example.

\section{Investigation of the linear response and comparison of the response relations}
\label{sec:Investigation}

\subsection{Setup and simulation parameters}
\label{subsec:Setup}
\begin{figure}[!t]
\includegraphics[width=0.8\linewidth]{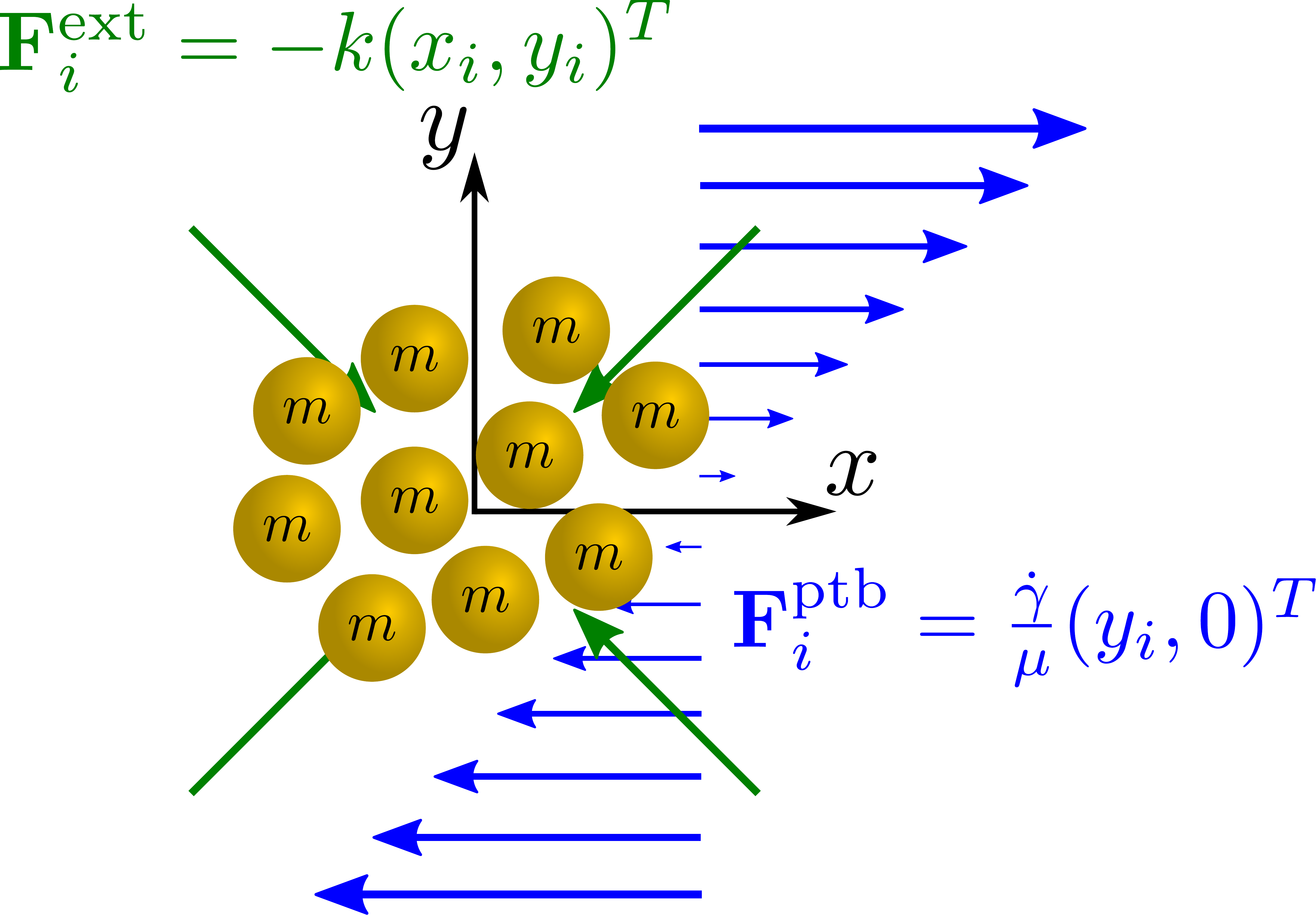}
	\caption{\label{fig:systemHP}Interacting Brownian particles (with a finite mass $ m $) confined in a harmonic potential and perturbed by simple shear flow in two space dimensions.}
\end{figure}

\begin{figure*}[!t]
\begin{tabular}{cc}
\includegraphics[width=0.49\linewidth]{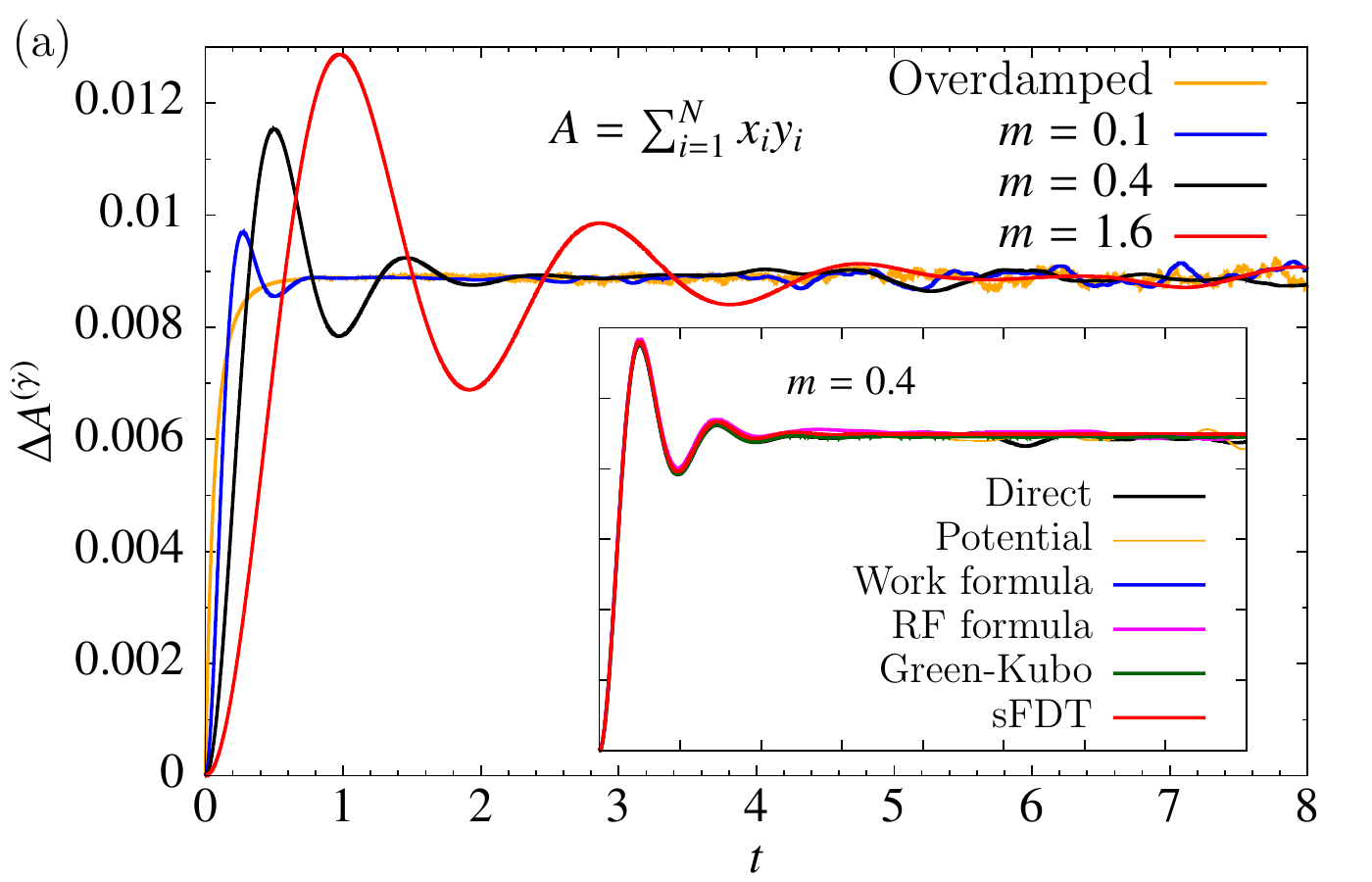}
&
\includegraphics[width=0.49\linewidth]{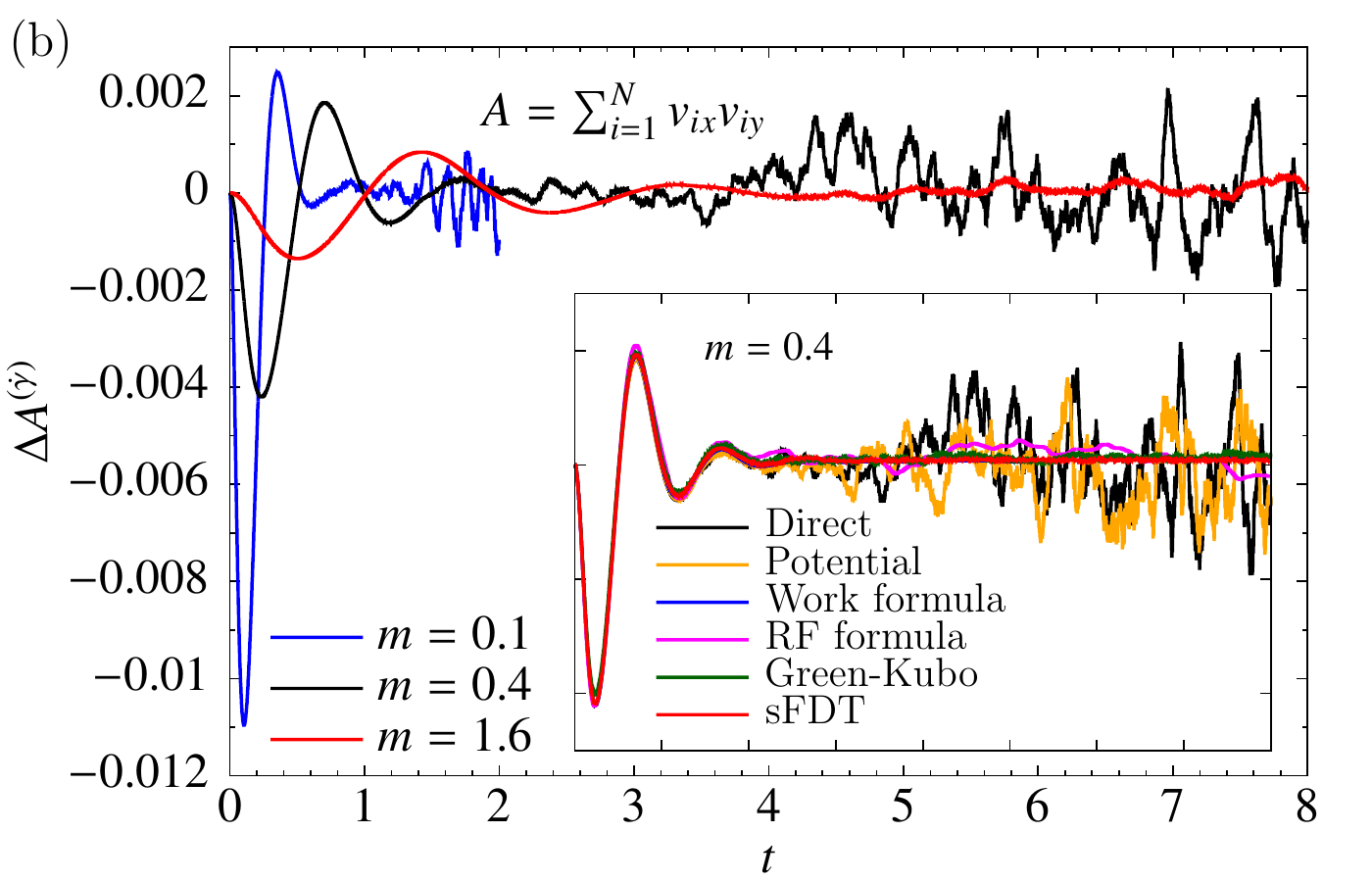}\\
\includegraphics[width=0.49\linewidth]{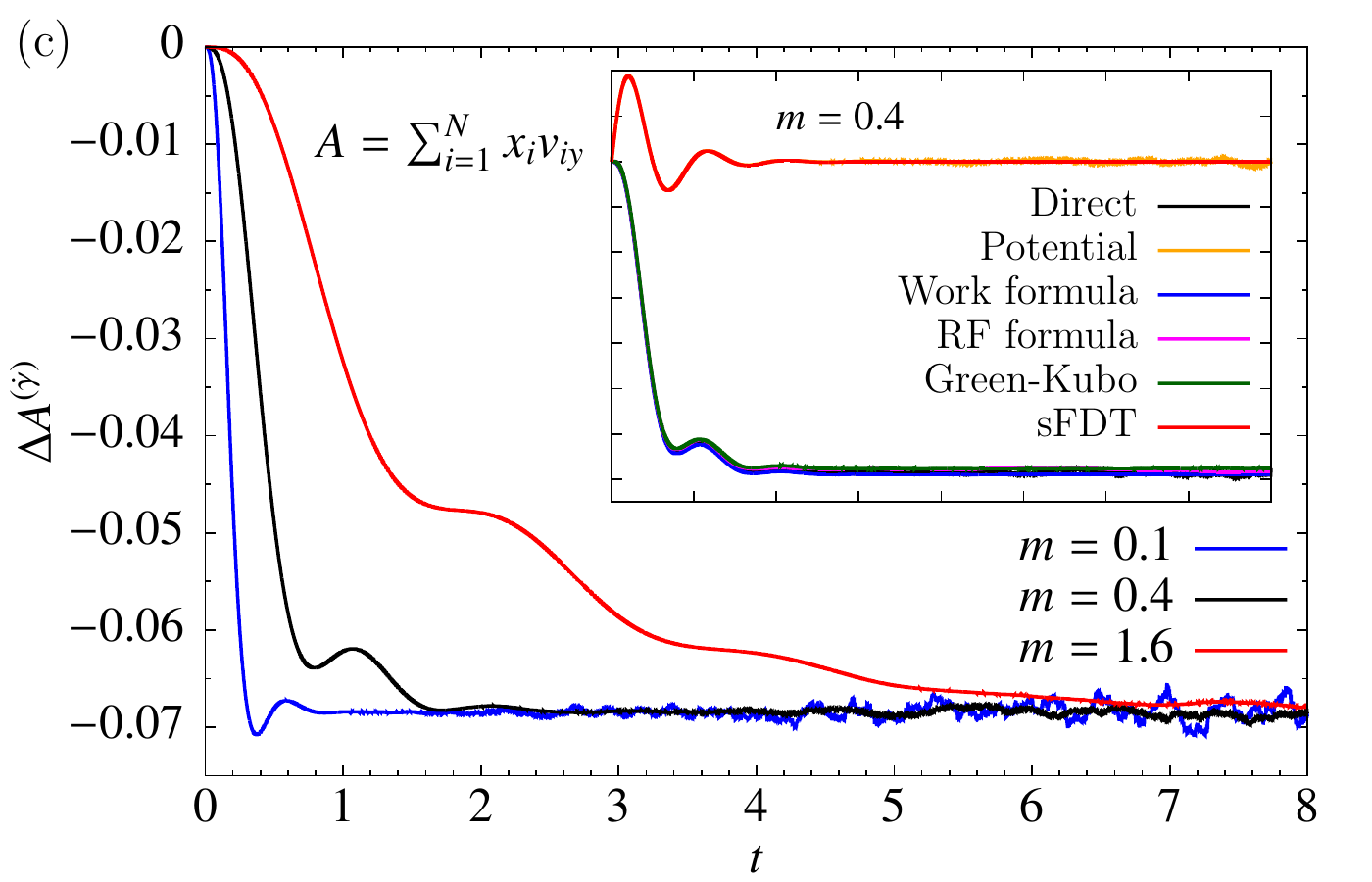}
&
\includegraphics[width=0.49\linewidth]{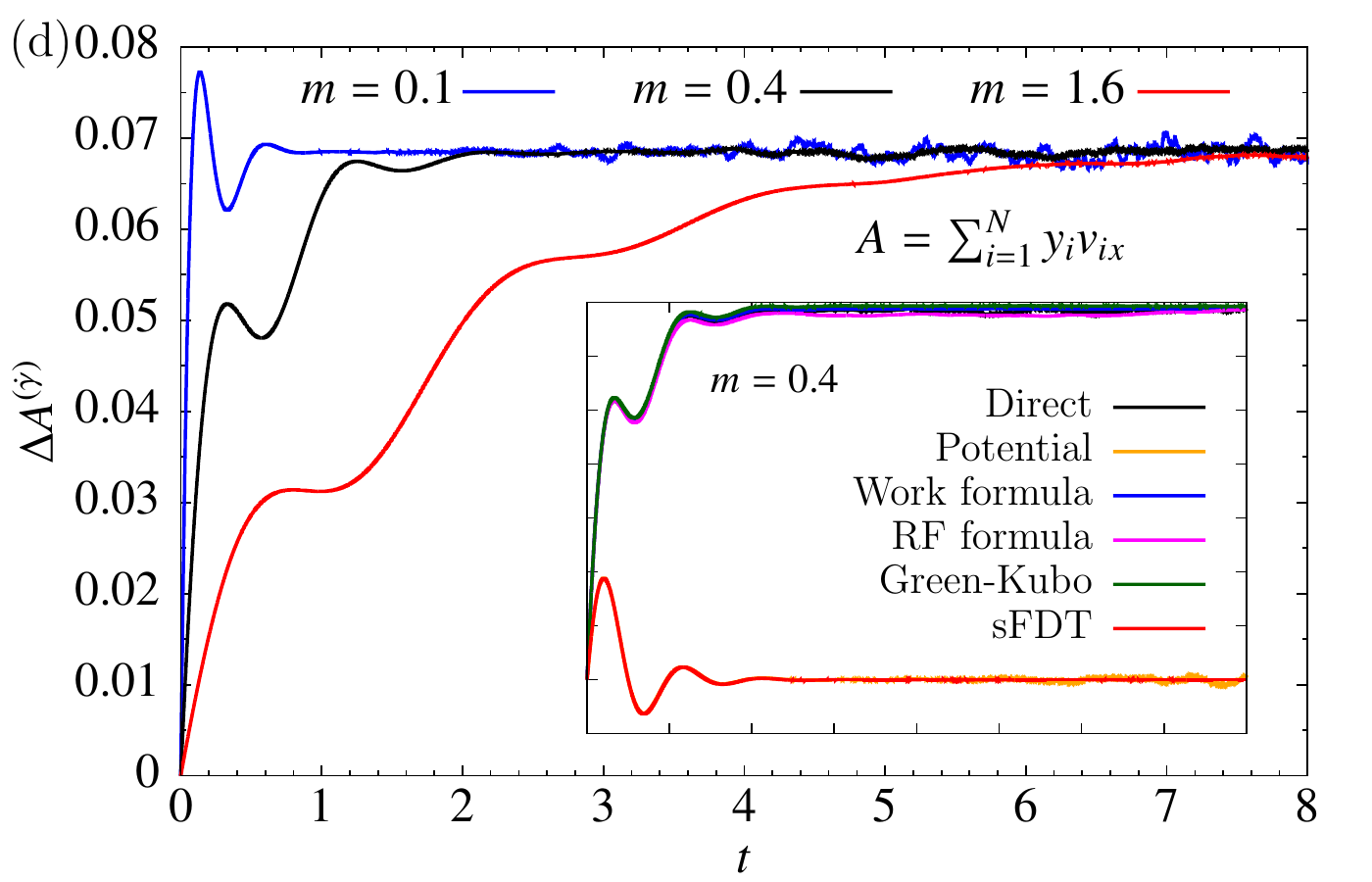}
\end{tabular}
\caption{\label{fig:Transient}Main plots show the directly computed linear response as a function of time after start of shear for the system depicted in Fig.~\ref{fig:systemHP}. Four different observables are considered. For each observable, the results for three different masses are shown; for $ A = \sum_{i=1}^Nx_iy_i $, the overdamped ($ m = 0 $) response is also plotted. The inset plots compare the linear response for $ m = 0.4 $ computed via the six ways introduced in Sec.~\ref{sec:Six_ways}: (i) directly perturbing by shear (the same as in the main plots, labeled \enquote{Direct}); (ii) perturbing by the shear potential~\eqref{eq:Ushear} [labeled \enquote{Potential}]; (iii -- vi) using linear response formulas~\eqref{eq:LR_TA},~\eqref{eq:LR_RF},~\eqref{eq:LR_GKU}, and~\eqref{eq:LR_RK} [labeled \enquote{Work formula}, \enquote{RF (random force) formula}, \enquote{Green-Kubo}, and \enquote{sFDT}, respectively]. Parameters: $ N = 10 $, $ \dot{\gamma} = 0.01 $, and $ C = 10^8 $.}
\end{figure*}

As a specific example, we study the response to shear of a two-dimensional system of interacting Brownian particles confined in a harmonic potential (see Fig.~\ref{fig:systemHP}). A similar scenario was studied in Ref.~\cite{Asheichyk2019_2}; here, we extend that study by (i) considering underdamped  dynamics and (ii) computing the response using Eqs.~\eqref{eq:ShearIsPotential},~\eqref{eq:LR_TA}, and~\eqref{eq:LR_RF}.

\begin{figure*}[!t]
\begin{tabular}{cc}
\includegraphics[width=0.5\linewidth]{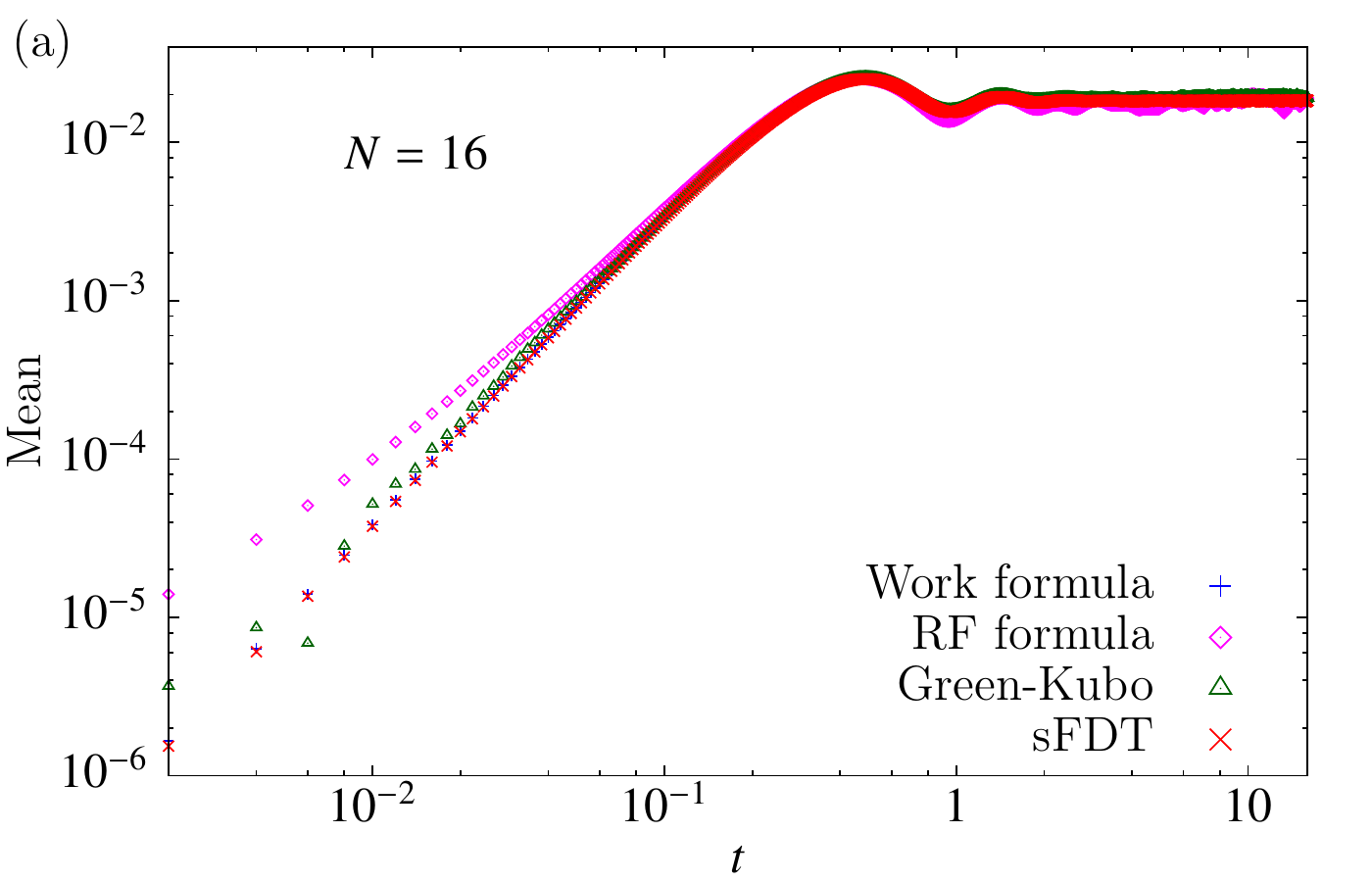}
&
\includegraphics[width=0.5\linewidth]{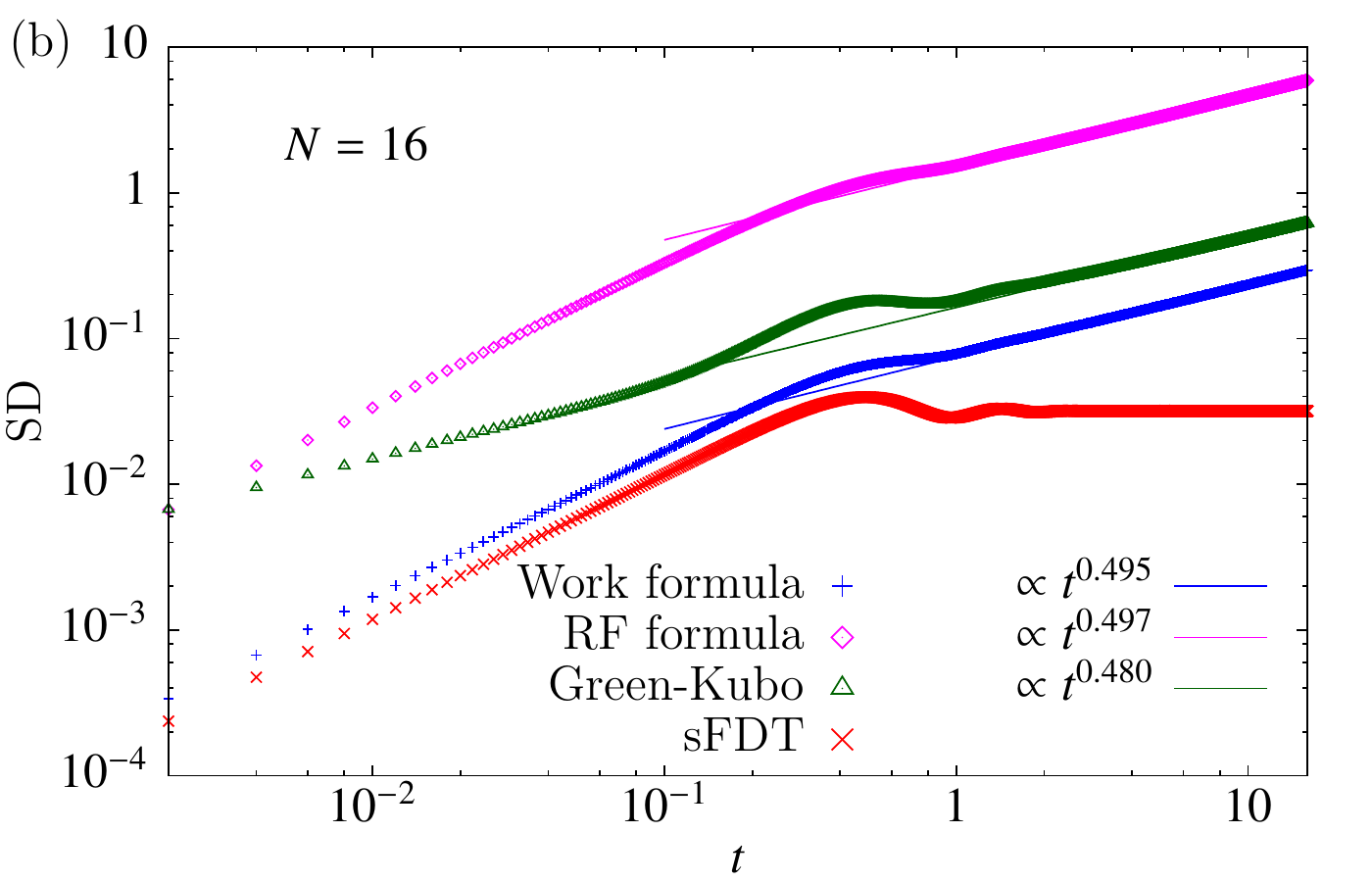}
\end{tabular}
\begin{tabular}{ccc}
\includegraphics[width=0.33\linewidth]{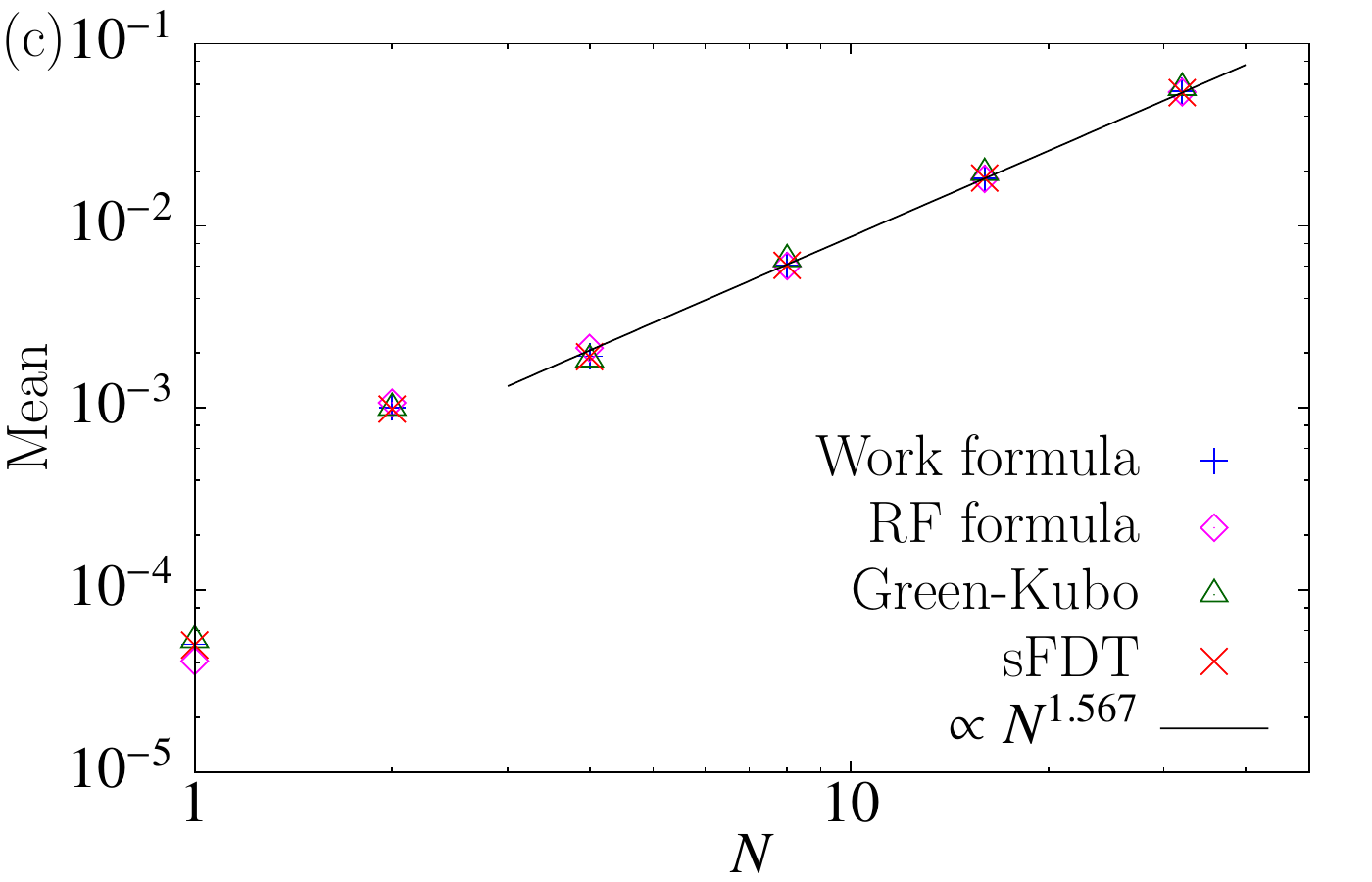}
&
\includegraphics[width=0.33\linewidth]{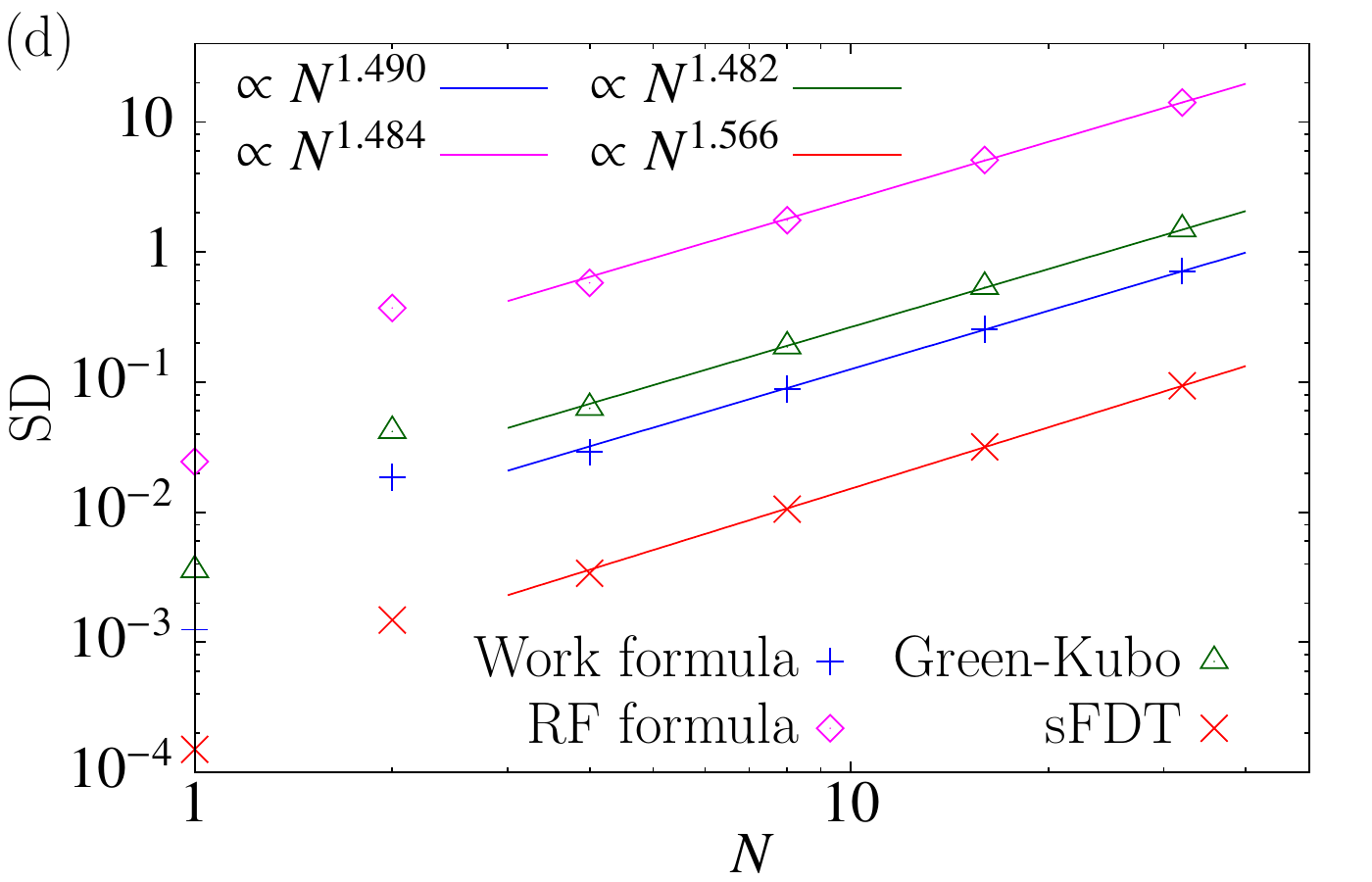}
&
\includegraphics[width=0.33\linewidth]{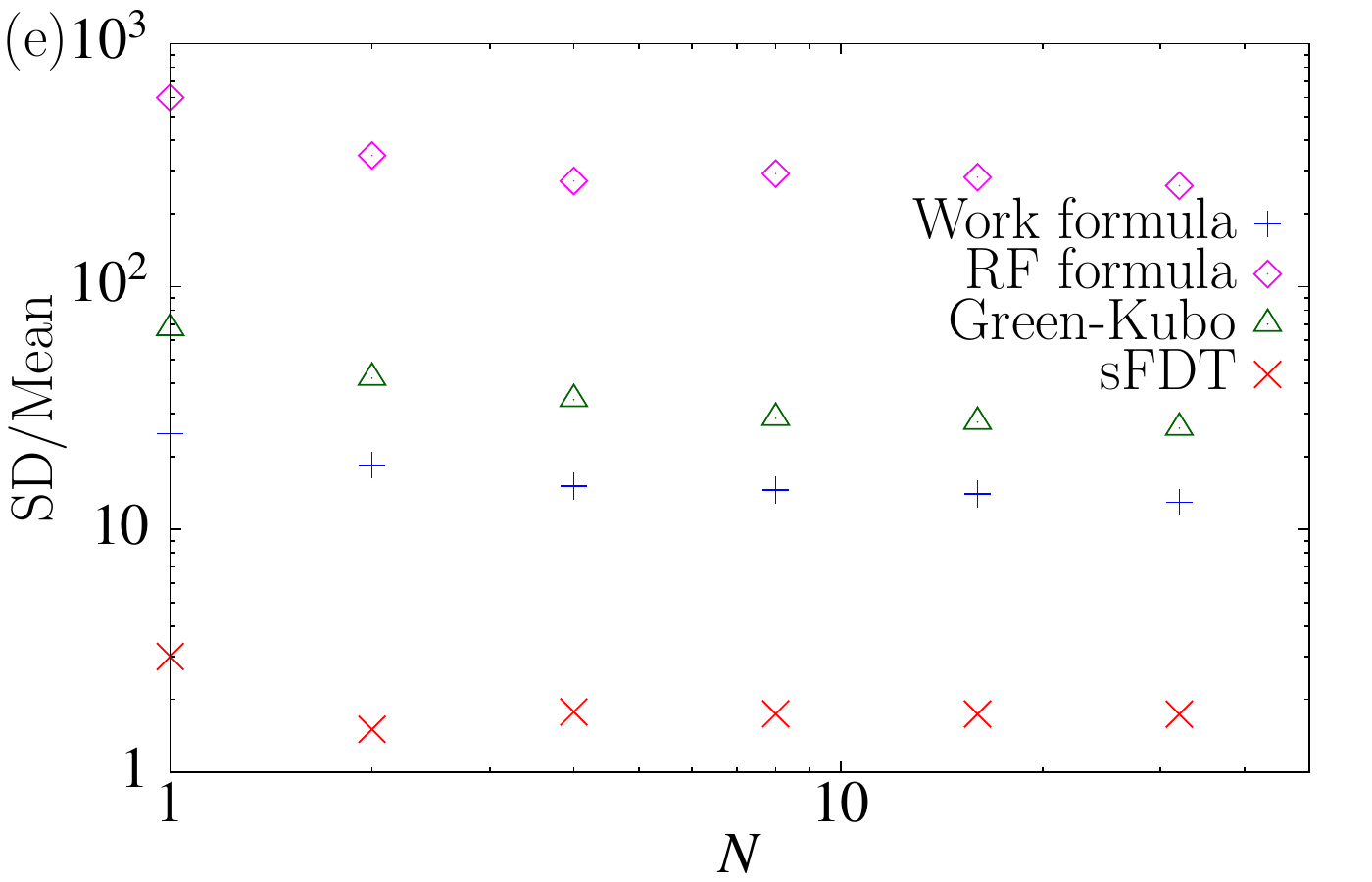}
\end{tabular}
\caption{\label{fig:Var}Panels~(a) and~(b) show the linear response (mean) and its standard deviation (SD) as functions of time  for the system depicted in Fig.~\ref{fig:systemHP} with $ N = 16 $ particles, computed using linear response formulas~\eqref{eq:LR_TA},~\eqref{eq:LR_RF},~\eqref{eq:LR_GKU}, and~\eqref{eq:LR_RK} [labeled \enquote{Work formula}, \enquote{RF (random force) formula}, \enquote{Green-Kubo}, and \enquote{sFDT}, respectively]. Panels~(c),~(d), and~(e) show the dependence of the stationary linear response (mean), its SD, and the relative SD (SD divided by the mean) on $ N $, also computed with the four formulas. Straight lines correspond to power-law fits. The considered observable is $ A = \sum_{i=1}^Nx_iy_i $. Parameters: $ m = 0.4 $, $ \dot{\gamma} = 0.01 $, and $ C = 10^6 $  (in addition, averaging over time in the steady state is performed; for details regarding the time average of SD, see the main text).}
\end{figure*}

The particles interact via a screened Coulomb potential
\begin{equation}
U^{\textrm{int}} = \frac{J}{2}\sum_{i=1}^N\sum_{j=1(j\neq i)}^N\frac{1}{r_{ij}} e^{-\frac{r_{ij}}{r^{\textrm{int}}}},
\label{eq:Uint}
\end{equation}
where $ r_{ij}\equiv|\vct{r}_i-\vct{r}_j|$, $J$, and $r^{\textrm{int}}$ are the interparticle distance, interaction strength, and interaction range, respectively. The external harmonic potential confining the particles reads
\begin{equation}
U^{\textrm{ext}} = \frac{k}{2}\sum_{i=1}^N|\vct{r}_i|^2.
\label{eq:Uext}
\end{equation}
For $N=1$, i.e.,  a single particle, we provide the analytical solution in Appendix~\ref{app:resp1},  complementing the results of Refs.~\cite{Rzehak2003, Holzer2010, Lander2012, Kahlert2012}  with the transient response. For $N>1$  particles, we evaluate the methods numerically, where we set $ k_{\textrm{B}}T = r^{\textrm{int}} = \mu = 1 $, $ J = 25 $, and $ k = 10 $; $ N $, $ m $, $ \dot{\gamma} $, and the number $ C $ of the independent noise realizations are varied between measurements. The dynamics is simulated using the Euler method with the time step $ \Delta t  = 5 \times 10^{-4} $ (note that in Figs.~\ref{fig:Var}(a),~\ref{fig:Var}(b), and~\ref{fig:respUptb} only every fourth time point is plotted).

\subsection{Transient response}
\label{subsec:Transient}
First, we study how the response of the system to shear depends on time. Interactions and a finite inertia give rise to interesting relaxation effects, as shown in Fig.~\ref{fig:Transient}. We consider observables $ A = \sum_{i=1}^Nx_iy_i $,  $ A = \sum_{i=1}^Nv_{ix}v_{iy} $, $ A = \sum_{i=1}^Nx_iv_{iy} $, and  $ A = \sum_{i=1}^Ny_iv_{ix} $, where the first two are $ xy $ symmetric, while the latter two are not. The shear rate $ \dot{\gamma} = 0.01 $ is chosen to be sufficiently small compared to other system parameters, such that the regime linear  in $ \dot{\gamma} $ applies.

The response for $ A = \sum_{i=1}^Nx_iy_i $ is given in Fig.~\ref{fig:Transient}(a). This quantity characterizes the morphology of the system: When shear is applied, the spatial distribution of the particles changes from circular to ellipsoidal (where principal axes of the ellipse are rotated with respect to the coordinate system), such that $ A $ becomes positive on average~\cite{Asheichyk2019_2, Rzehak2003, Holzer2010, Lander2012, Kahlert2012}. For overdamped particles, there is a simple monotonic relaxation towards the steady state, as was observed in Ref.~\cite{Asheichyk2019_2}. The situation is different if the particles have a finite mass, where oscillations appear. These oscillations become more pronounced as the mass increases: They have a larger amplitude, as well as a longer period and relaxation time. It is worth noticing that such an interacting system behaves qualitatively different compared to a single-particle case [see Fig.~\ref{fig:AE}(a)]. While the response of the former exceeds the steady-state result during oscillations, the response of the latter does not (here, the maxima of oscillations lie exactly on the steady-state result). The difference may be attributed to elastic energy stored in particle interactions, which leads to stronger oscillations. 
  
Figure~\ref{fig:Transient}(b) shows the response for $ A = \sum_{i=1}^Nv_{ix}v_{iy} $. Here, the period of oscillations and the relaxation time increase with the mass, but the amplitude decreases. The response approaches zero for large times, as was also observed for a single particle (see Fig.~\ref{fig:AE}(b) and Refs.~\cite{Holzer2010, Kahlert2012}). However,  the transient response can be positive for a many-particle system, whereas it is always below or equals zero for a single particle [compare Figs.~\ref{fig:Transient}(b) and~\ref{fig:AE}(b)].

The responses for $ A = \sum_{i=1}^Nx_iv_{iy} $ and $ A = \sum_{i=1}^Ny_iv_{ix} $ are shown in Figs.~\ref{fig:Transient}(c) and~\ref{fig:Transient}(d), respectively. The former is negative, while the latter is positive, and their steady-state values are related by a minus sign. However, this is not true for all times, as can be clearly seen from the peaks for $ m = 0.1 $ and $ m = 0.4 $, as well as from the behavior at $ t = 0 $ (the curves have zero derivatives for $ A = \sum_{i=1}^Nx_iv_{iy} $, but finite ones for $ A = \sum_{i=1}^Ny_iv_{ix} $). The reason for the difference originates in the different roles played by the velocity (viz.~$x$) and the gradient (viz.~$y$) direction. Also for these observables, one can identify qualitative differences compared to single-particle responses [see Figs.~\ref{fig:AE}(c) and~\ref{fig:AE}(d)]: For example, the curve for $ m = 0.1 $ in Fig.~\ref{fig:Transient}(c) goes below the steady-state result as the response develops in time, which is not observed for the curves in Fig.~\ref{fig:AE}(c). 

It is interesting that the steady-state linear response does not depend on the mass, as can be seen in Fig.~\ref{fig:Transient}. In general, the mass dependence is expected even for spatial observables like $ A = \sum_{i=1}^Nx_iy_i $, because shear drives the system out of equilibrium leading to the coupling between spatial and velocity distributions. However, this effect appears only in a nonlinear order, as shown in Sec.~\ref{sec:Nonlinear}. We further note that, for $ A = \sum_{i=1}^Nx_iy_i $ and $ A = \sum_{i=1}^Nv_{ix}v_{iy} $, the mass independence is an exact and direct consequence from sFDT~\eqref{eq:LR_RK}: In the steady-state limit, the last term in Eq.~\eqref{eq:LR_RK} vanishes; the first one, being an equilibrium average, does not depend on the mass if $ A $ contains only particle positions, and vanishes if $ A = \sum_{i=1}^Nv_{ix}v_{iy} $ [confirming the zero result for large times in Fig.~\ref{fig:Transient}(b)]. This demonstrates another advantage of sFDT~\eqref{eq:LR_RK} compared to other response relations: Thanks to its simple structure, sFDT allows to make important statements without computations.%, which appears more difficult when using other formulas.

The inset plots of Fig.~\ref{fig:Transient} show the response for $ m = 0.4 $ computed via the six ways introduced in Sec.~\ref{sec:Six_ways}. For $ A = \sum_{i=1}^Nx_iy_i $ and $ A = \sum_{i=1}^Nv_{ix}v_{iy} $ all the ways agree; this demonstrates their validity. For $ A = \sum_{i=1}^Nx_iv_{iy} $ and $ A = \sum_{i=1}^Ny_iv_{ix} $, the response to the shear potential~\eqref{eq:Ushear} and the response computed using sFDT~\eqref{eq:LR_RK} disagree from the other ways, as expected for not $ xy $-symmetric observables.

\subsection{Standard deviation}
\label{subsec:Variance}
Our next investigation concerns the standard deviation (SD) of linear response relations. According to the central limit theorem, an average has a statistical error $ \frac{\sigma}{\sqrt{C}} $, where $ \sigma $ (not to be confused with the stress tensor $ \sigma_{xy} $) is SD of the underlying distribution, and $ C $ is the number of independent measurements. A larger $ \sigma $ hence requires a larger $ C $ (and thus more computational resources) to obtain a given accuracy.  

The response formulas introduced in Sec.~\ref{sec:Six_ways}, where the averaged quantity $ B $ is different, give the same average, or mean, value of $ B $. For example, in formula~\eqref{eq:LR_TA}, $ B =  \frac{\dot{\gamma}}{k_{\textrm{B}}T\mu}\int_0^tdt'A(t)\sum_{i=1}^Ny_i(t')v_{ix}(t') $, while in formula~\eqref{eq:LR_RK}, $ B = \frac{\dot{\gamma}}{2k_{\textrm{B}}T\mu}\left[A\sum_{i=1}^Nx_iy_i - A(t)\sum_{i=1}^Nx_i(0)y_i(0)\right] $. However, SD, defined  as $ \sigma = \sqrt{\left\langle B^2\right\rangle - {\left\langle B\right\rangle}^2} $, is, in general, different for different formulas, as we demonstrate below. Therefore, the practical usage of each formula is based not only on the quantity $ B $ which has to be measured, but also on SD of the formula. Since SD tells us about the statistical efficiency of the computed average, it is a crucial property of linear response formulas: The smaller $ \sigma $, the more efficient the formula.

In Ref.~\cite{Asheichyk2019_2}, we computed SD for formulas~\eqref{eq:LR_GK} and~\eqref{eq:LR_RK} using overdamped Brownian particles. Here, we compute SD using the underdamped system depicted in Fig.~\ref{fig:systemHP}, consider more response relations, and study the transient regime. The results are given in Fig.~\ref{fig:Var}, with $ A = \sum_{i=1}^Nx_iy_i $.

Figure~\ref{fig:Var}(a) shows the transient linear response to shear for $ N = 16 $ particles computed using formulas~\eqref{eq:LR_TA},~\eqref{eq:LR_RF},~\eqref{eq:LR_GKU}, and~\eqref{eq:LR_RK}. As expected, all four curves agree and relax to a definite stationary value. In contrast, the corresponding SD, plotted in Fig.~\ref{fig:Var}(b), is different for each formula. For sFDT, SD converges to a steady-state value, while it diverges with $t$ for the other formulas. For a single overdamped particle, we obtained a divergence with $ \sqrt{t} $ analytically (see Appendix~\ref{app:var}), and the numerical data in Fig.~\ref{fig:Var}(b) also suggest such law for the case of $N>1$. When $N>1$, this divergence is also analytically suggested, and it is  attributed to the fact that SD of these formulas involves double integration over time.

The absence of a genuine steady-state value for SD of formulas~\eqref{eq:LR_TA},~\eqref{eq:LR_RF},~\eqref{eq:LR_GKU} raises a question about a usefulness of a comparison. However, there is no fundamental problem, as, indeed, the statistical error of a numerically computed average accumulates with time due to time integration. A natural choice for the steady-state SD is then based on the steady-state mean. Here, the steady-state value for the mean is reached at $ t \approx 8 $ for all considered $ N $ [see Fig.~\ref{fig:Var}(a) for $ N = 16 $], and we estimate the steady-state value for SD at the same time, i.e.,  at $ t \approx 8 $. In order to perform time averaging in the steady state, we estimate the stationary mean and SD as an average over time in the interval $ t \in [8, 16] $.

Lower panels of Fig.~\ref{fig:Var} show the dependence of the stationary mean, SD, and the relative SD on the number of particles $ N $. For the mean, the four response relations agree and scale as $ N^{1.567} $ (for $ N \gtrapprox 4 $). However, SD is very different for each response formula [as can also be seen from the transient SD in Fig.~\ref{fig:Var}(b)], although the scaling behavior is similar. Formula~\eqref{eq:LR_RK}, sFDT, has the lowest SD, while formula~\eqref{eq:LR_RF} containing the random force -- the highest one. Estimated from SD and the central limit theorem, Eq.~\eqref{eq:LR_RK} thus needs roughly a factor of $ 10^4 $ smaller number of independent simulation runs than Eq.~\eqref{eq:LR_RF} for the same statistical error (for example, this is a minute of computational time versus 7 days). The efficiencies of the work formula~\eqref{eq:LR_TA} and the Green-Kubo relation~\eqref{eq:LR_GKU} are in between, respectively. Their SDs differ between each other roughly by a factor of $ 2 $, but they are separated from the curves for Eqs.~\eqref{eq:LR_RK} and~\eqref{eq:LR_RF} by an order of magnitude. Note that the relative SD, divided by the mean, is independent of $ N $ (for $ N \gtrapprox 4 $) for all response relations, as can be seen in Fig.~\ref{fig:Var}(e). This is different from the typical behavior of $1/\sqrt{N}$, as the system is not extensive with $N$. 

%The found relative for the variance my depend on the type of system studied.

We also would like to mention that, in contrast to the mean, SD computed for formulas~\eqref{eq:LR_TA} and~\eqref{eq:LR_RF} can depend on the discretization scheme used in numerical simulations, i.e., Ito or Stratonovich \cite{Asheichyk2020, Cugliandolo2017, Gardiner2010, Wynants2010}.

\section{Linear versus nonlinear}
\label{sec:Nonlinear}
Finally, we would like to demonstrate some differences between the linear and nonlinear responses to shear. 

It was observed that the relaxation to the steady state is very different between the two cases~\cite{Asheichyk2020, Asheichyk2019_2, Amann2013}. What about the steady state itself? One important question is whether the steady-state response depends on the mass, and if it does,  for what observables. In the linear order, it does not, as can be seen in Fig.~\ref{fig:Transient} [see also discussions in Subsec.~\ref{subsec:Transient}]. 

The situation is different for the nonlinear case. Table~\ref{table:NL_mass} compares how the response for $ \dot{\gamma} = 0.01 $ and $ \dot{\gamma} = 20 $ depends on the mass, using the observable $ A = \sum_{i=1}^Nx_iy_i $, notably a function of position. When $ \dot{\gamma} = 0.01 $ (linear response), there is no dependence (up to statistical error), as we just mentioned. When $ \dot{\gamma} = 20 $, however, a clear dependence can be seen. This shows that the steady-state distribution couples inertia and potential parts. The dependence of the spatial distribution on the particle mass was very recently observed for a single Brownian gyrator~\cite{Bae2020}, which is also a nonequilibrium system.

\begin{table}[!t]
\caption{\label{table:NL_mass}The directly computed steady-state response to shear for different shear rates and masses. The system is the same as in Sec.~\ref{sec:Investigation}; $ A = \sum_{i=1}^Nx_iy_i $. Parameters: $ N = 10 $ and $ C = 10^6 $ (in addition, averaging over time in the steady state is performed).}
\begin{ruledtabular}
\begin{tabular}{c|ccc}
\diagbox{$\dot{\gamma}$}{$m$} & $0$ & $0.1$ & $0.4$\\
\hline
$0.01$ & $0.00881$ & $0.00877$ & $0.00923$\\
$20$ & $10.27$ & $14.71$ & $31.02$
\end{tabular}
\end{ruledtabular}
\end{table} 

\begin{figure}[!b]
\includegraphics[width=1.0\linewidth]{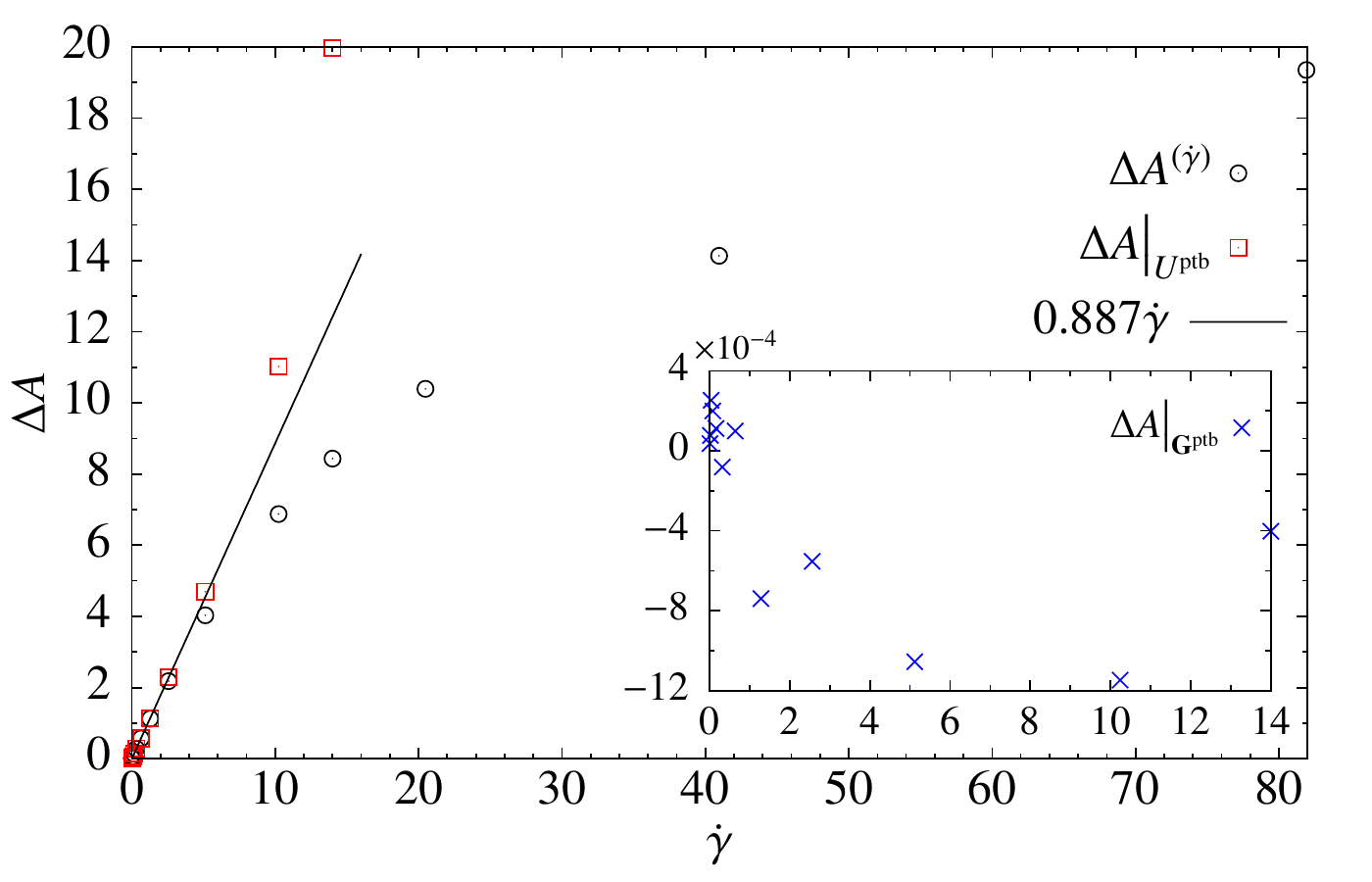}
	\caption{\label{fig:NL_GD_lin}Overdamped steady-state responses to shear ($ \Delta A^{(\dot{\gamma})} $), the shear potential ($ \Delta A\big|_{U^{\textrm{ptb}}} $), and the rotation force ($ \Delta A\big|_{\vct{G}^{\textrm{ptb}}} $, the inset plot) as functions of the shear rate $ \dot{\gamma} $. The system is the same as in Sec.~\ref{sec:Investigation}; $ A = \sum_{i=1}^Nx_iy_i $. The black line represents the linear fit using the five smallest values of $ \Delta A^{(\dot{\gamma})} $. Parameters: $ N = 10 $ and $ C = 10^6 $ (in addition, averaging over time in the steady state is performed). Note that $ \Delta A\big|_{\vct{G}^{\textrm{ptb}}} =0$ within accuracy.}
\end{figure}

Another important question to investigate is whether our main observation, Eq.~\eqref{eq:ShearIsPotential}, remains valid beyond the linear order. Remember that Eq.~\eqref{eq:ShearIsPotential} follows from two facts: (i) the superposition principle [Eq.~\eqref{eq:SPP}] and (ii) the fact that the response to the rotation force $ \vct{G}_i^{\textrm{ptb}} $ is zero for $xy$-symmetric systems and observables. 

Figure~\ref{fig:NL_GD_lin}, showing the dependence of $ \Delta A^{(\dot{\gamma})} $, $ \Delta A\big|_{U^{\textrm{ptb}}} $, and  $ \Delta A\big|_{\vct{G}^{\textrm{ptb}}} $ on $ \dot{\gamma} $, answers this question. The system and observable considered in the figure are $xy$-symmetric. One can see that $ \Delta A^{(\dot{\gamma})} = \Delta A\big|_{U^{\textrm{ptb}}}$ up to $ \dot{\gamma} \approx 1 $, where the response to shear is linear in $ \dot{\gamma} $. Once $ \Delta A^{(\dot{\gamma})} $ deviates from the linear behavior (it can be already seen for $ \dot{\gamma} \approx 2.5 $), Eq.~\eqref{eq:ShearIsPotential} fails. Since, for symmetry reasons, $ \Delta A\big|_{\vct{G}^{\textrm{ptb}}} $ remains zero even when this deviation occurs, Eq.~\eqref{eq:ShearIsPotential} fails only because the superposition principle~\eqref{eq:SPP} fails. We note, however, that for other $xy$-symmetric observables, e.g. $ A = \sum_{i=1}^N(x_i^2+y_i^2) $, the nonlinear response to $ \vct{G}_i^{\textrm{ptb}} $ is not necessarily zero, because the effect of $ \vct{G}_i^{\textrm{ptb}} $ for large $ \dot{\gamma} $ may be different from a pure rotation.

It is also worth noticing that the system perturbed by $ U^{\textrm{ptb}} $ has a transition at $ \dot{\gamma} = 2\mu k $ (see Appendix~\ref{app:respUptb}). For  $ \dot{\gamma} \geq 2\mu k $ (which corresponds to $ \dot{\gamma} \geq 20 $ for the chosen parameters), the system is unstable, because the sum of $ U^{\textrm{ptb}} $ and $ U^{\textrm{ext}} $ form a saddle potential, see Fig.~\ref{fig:respUptb}. The response thus diverges with time for $ \dot{\gamma} \geq 2\mu k $, and the steady-state response diverges when approaching $ \dot{\gamma} = 2\mu k $ from below. The points for $ \Delta A\big|_{U^{\textrm{ptb}}}$ and $ \dot{\gamma} \geq 2\mu k $ are absent in Fig.~\ref{fig:NL_GD_lin}.

\section{Conclusion}
\label{sec:Conclusion}
In this paper, we studied the linear response of interacting Brownian particles with a finite mass to simple shear flow. The superposition principle of the linear response [Eq.~\eqref{eq:SPP}], and the symmetry of the system and the observable with respect to the interchange of the $ x $ and $ y $ axes (shear axes)  make the responses to shear and to the shear potential [Eq.~\eqref{eq:Ushear}] equivalent. This fact leads to two efficient ways for the linear response computation: perturbation by the shear potential in Eq.~\eqref{eq:Ushear}, and FDT for shear [sFDT, Eq.~\eqref{eq:LR_RK}].

Using six different computational methods (direct perturbation by shear, perturbation by the shear potential, and applying four linear response relations), we computed the transient response to shear for four different observables. It was found that interparticle interactions and a finite inertia lead to nontrivial time dependence of the response, which is very different compared to a single underdamped particle or interacting overdamped particles. The four response relations were compared in terms of their statistical efficiency, where sFDT was found to be the most efficient relation, requiring about 100 times less computational resources than the next efficient one. This is to be expected due to the absence of a time integral, which also simplifies the computation in other regards. Equation~\eqref{eq:LR_RK}, for the steady state, can even be evaluated in a simple Monte Carlo simulation, in contrast to other formulas, where the dynamics needs to be resolved for the time integration.

Going beyond the linear order, we observed that, in contrast to the linear response, the steady-state nonlinear response depends on the particle mass. Also, for large shear rates, the response to shear cannot be computed via the response to the shear potential, because the superposition principle fails.

Future work may study the response to shear of active Brownian particles with a finite mass~\cite{Lowen2020}. Application of the discussed here computational methods to the viscosity of Brownian suspensions is also a promising avenue to explore. Another interesting point is the connection between the response relations discussed here and those following from Newtonian dynamics (where DOLLS and SLLOD equations are used to model shear)~\cite{Evans2008, Chong2009, Suzuki2013}.

\begin{acknowledgments}
K.A. was supported by the German Academic Scholarship Foundation (Studienstiftung des deutschen Volkes) and by the Physics Department of the University of Stuttgart. M.F. and M.K. acknowledge funding by the DFG within SFB 1432, through projects C06 and C05, respectively. K.A. also acknowledges support by S.~\mbox{Dietrich} and Alexander N. Furs.
\end{acknowledgments}

% MAIN_PART_end --------------------------------------------

% FOOTER_begin ---------------------------------------------
\begin{appendix}
\section{Linear response in Klein-Kramers equation}
\label{app:KKE}
The Langevin equation~\eqref{eq:LE} corresponds to the Klein-Kramers equation $\partial_t \Psi = \Omega \Psi$ for the probability distribution function  (pdf) $\Psi$ given by the Fokker-Planck operator~\cite{Hess1983}
\begin{eqnarray}\label{eq1}
\notag  \Omega &=& \Omega_0+\delta\Omega= - \sum_{i=1}^N\left(\vct{F}_i \cdot \frac{\partial}{  \partial m \vct{v}_i}+ \vct{v}_i  \cdot \frac{\partial}{\partial \vct{r}_i}\right)   \\
                &+&
                 \frac{1}{\mu } \sum_{i=1}^N   \frac{\partial}{\partial m \vct{v}_i} \cdot\left( k_{\textrm{B}} T \frac{\partial}{\partial m\vct{v}_i}+ \vct{v}_i- \boldsymbol{\kappa}\cdot \vct{r}_i \right),
\end{eqnarray}
where $\vct{F}_i$ denotes all potential forces acting on particle $i$ and the flow rate tensor $\boldsymbol{\kappa}$ has nonvanishing element $\kappa_{xy}=\dot\gamma\Theta(t)$, where $\Theta(t)$ is the Heaviside step function. Note in passing that the perturbation $\delta \Omega=- \frac{1}{\mu}\sum_{i=1}^N \frac{\partial}{\partial m\vct{v}_i}\cdot  \boldsymbol{\kappa}\cdot \vct{r}_i $ arises in the diffusion term of the Fokker-Planck operator and not in its drift term where a coupling to an external flow field would enter in Liouvillian response~\cite{Kubo1966}. 

To linear order, the change of the pdf obeys the differential equation 
\begin{equation}\label{eq2} 
\left(\partial_t - \Omega_0 \right) \; \delta \Psi = \delta \Omega\;  \Psi_{\rm eq} = \frac{1}{k_{\textrm{B}}T} \; \Pi \; \Psi_{\rm eq},
\end{equation}
 where $\Psi_{\rm eq}$ is the canonical equilibrium pdf and the inhomogeneity is produced by the power $\Pi$~\cite{Kurchan1998}
\begin{equation}\label{eq3}  
\Pi= \frac{1}{\mu}\; \sum_{i=1}^N\;  \vct{v}_i \cdot \boldsymbol{\kappa}\cdot \vct{r}_i.
\end{equation}
For the specified constant perturbation starting at $t=0$, the change of the pdf becomes
\begin{equation}\label{eq4} 
 \delta \Psi(t) = \frac{1}{k_{\textrm{B}}T}  \int_0^t dt'\;  e^{ \Omega_0(t-t')}\;  \Pi\; \Psi_{\rm eq}.
\end{equation}
This leads to Eq.~\eqref{eq:LR_TA}. For an arbitrary perturbation $ \vct{K}^{\textrm{ptb}}_i $, the power in Eq.~\eqref{eq3} is replaced by $ \Pi= \sum_{i=1}^N\;  \vct{v}_i \cdot \vct{K}^{\textrm{ptb}}_i $, leading to Eq.~\eqref{eq:G_LR} [with $ \vct{K}^{\textrm{ptb}}_i =  \vct{G}^{\textrm{ptb}}_i] $.

In the general case, the condition of detailed balance, $\Omega_0(\{\vct{r}_i, \vct{v}_i\})\,A\, \Psi_{\rm eq} = \Psi_{\rm eq}\, \Omega^\dagger_0(\{\vct{r}_i, -\vct{v}_i\})\,A$ with $\Omega^\dagger_0$ the adjoint and $A$ an arbitrary function~\cite{Risken1996}, can be used to show that
\begin{equation}\label{eq5} 
  \Pi\; \Psi_{\rm eq}  =  \boldsymbol{\sigma} : \boldsymbol{\kappa}\;  \Psi_{\rm eq}  - \mu m\;\Omega_0\;  \Pi\; \Psi_{\rm eq},
\end{equation}
where $\boldsymbol{\sigma}$ is the Irving-Kirkwood stress tensor~\cite{Hansen2009} whose $xy$ element is given in Eq.~\eqref{eq:sigma}.
Integrating Eq.~\eqref{eq4} for $\delta\Psi$ gives two terms in this case:
\begin{align}
\notag \delta \Psi(t) & = \frac{1}{k_\textrm{B}T} \int_0^tdt'\; e^{ \Omega_0 t'} \;\boldsymbol{\sigma} : \boldsymbol{\kappa}\;  \Psi_{\rm eq}\\
& + \frac{\mu m}{k_{\textrm{B}}T} \left(1- e^{ \Omega_0 t}\right) \; \Pi\; \Psi_{\rm eq}, 
\label{eq6} 
\end{align}
which directly leads to Eq.~\eqref{eq:LR_GKU}. 

In the case of $xy$ symmetry, the flow can be described by the potential, $ \frac{1}{2\mu}\left(\boldsymbol{\kappa}+\boldsymbol{\kappa}^T\right)\cdot\vct{r}_i = -\boldsymbol{\nabla}_iU^{\textrm{ptb}} $, as discussed in Sec.~\ref{sec:Symmetries}. This allows to rewrite $\Pi\Psi_{\rm eq} = \Omega_0  U^{\rm ptb} \Psi_{\rm eq}$ [where $ \Pi $ is given by Eq.~\eqref{eq3}, with $ \boldsymbol{\kappa} $ replaced by $ \frac{1}{2}(\boldsymbol{\kappa}+\boldsymbol{\kappa}^T)$, and $ U^{\textrm{ptb}} $ is given by Eq.~\eqref{eq:Ushear}], and to perform the time integration in Eq.~\eqref{eq4} leading to $ \delta \Psi(t) = \frac{-1}{k_\textrm{B}T} (1- e^{ \Omega_0 t}) \; U^{\rm ptb} \Psi_{\rm eq}$. Performing an average with this $\delta\Psi$ leads to Eq.~\eqref{eq:LR_RK}.

\section{The symmetry of an equilibrium time correlation function}
\label{app:symm_eqcorr}
An equilibrium time dependent correlation function of two variables $ A $ and $ B $ can be written as (we consider $ t \geq t' $)~\cite{Risken1996}
\begin{align}
\notag \left \langle A(t)B(t') \right\rangle = & \left \langle A(t-t')B(0) \right\rangle =\\
 & \int d\Gamma A(\Gamma)e^{\Omega_0(\Gamma)(t-t')}B(\Gamma)\Psi_{\textrm{eq}}(\Gamma),
\label{eq:EqCorr}
\end{align}
	where $ \Gamma $ is the phase space, and $ \Omega_0 $ and $ \Psi_{\textrm{eq}} $ are the equilibrium Fokker-Planck operator (see Eq.~\eqref{eq1} and Refs.~\cite{Risken1996, Hess1983}) and distribution function, respectively. For an $ xy $-symmetric system, $ \Omega_0 $ and $ \Psi_{\textrm{eq}} $ are $ xy $ symmetric. Because of this, making the interchange $ x_i \leftrightarrow y_i, \ v_{ix} \leftrightarrow v_{iy} $ for $ A $ and $ B $ in Eq.~\eqref{eq:EqCorr} is equivalent to simply renaming the integration variables in the same way. The latter action does not affect the value of the integral. Therefore, an equilibrium correlation function computed for an $ xy $-symmetric system is also $ xy $ symmetric.

\section{Derivation of the underdamped Green-Kubo relation}
\label{app:GKU_derivation}
We demonstrate here the derivation of Eq.~\eqref{eq:LR_GKU} \cite{Asheichyk2020}. It is based on a path integral formalism~\cite{Altland2010, Asheichyk2019_1, Asheichyk2020}.

The path weight $ \mathcal{W}^{(\dot{\gamma})} $, representing the probability that within a time interval $ [0, t] $ the system described by Eq.~\eqref{eq:LE} follows a certain trajectory, reads~\cite{Altland2010, Asheichyk2019_1, Asheichyk2020, Cugliandolo2017, Onsanger1953, Machlup1953, Martin1973, Janssen1976, DeDominicis1978}
\begin{equation}
\mathcal{W}^{(\dot{\gamma})} \propto e^{-\mathcal{A}^{(\dot{\gamma})}}, \ \ \ \ \mathcal{A}^{(\dot{\gamma})} = \frac{\mu}{4k_{\textrm{B}}T}\textrm{Ito}\int_0^tdt'\sum_{i=1}^N\boldsymbol{\mathcal{X}}_i^2(t', \dot{\gamma}),
\label{eq:PW}
\end{equation}
where $ \mathcal{A}^{(\dot{\gamma})} $ is the system action and $ \boldsymbol{\mathcal{X}}_i^2 $ measures the deviation from the noise-free path,
\begin{equation}
\boldsymbol{\mathcal{X}}_i(\dot{\gamma}) = m\dot{\vct{v}}_i - \frac{1}{\mu}\boldsymbol{\kappa}\cdot\vct{r}_i +\frac{1}{\mu}\vct{v}_i -  \vct{F}^{\textrm{int}}_i - \vct{F}^{\textrm{ext}}_i .
\label{eq:X}
\end{equation}
Since Eq.~\eqref{eq:PW} is derived using the Ito time-discretization scheme~\cite{Asheichyk2020, Cugliandolo2017, Gardiner2010, Wynants2010}, the integral in Eq.~\eqref{eq:PW} is understood as a stochastic Ito integral~\cite{Asheichyk2020, Cugliandolo2017, Gardiner2010, Wynants2010}, as indicated by the notation \enquote{Ito}.

Expanding the path weight~\eqref{eq:PW} in powers of $ \dot{\gamma} $, we get
\begin{equation}
\mathcal{W}^{(\dot{\gamma})} - \mathcal{W} = \mathcal{W}\frac{\dot{\gamma}}{2k_{\textrm{B}}T}\textrm{Ito}\int_0^tdt'\sum_{i=1}^N\mathcal{X}_{ix}(t')y_i(t') + \mathcal{O}(\dot{\gamma}^2),
\label{eq:PW_expansion_Ito}
\end{equation}
where $ \mathcal{X}_{ix} $ and $ \mathcal{W} $ correspond to the unperturbed system. One can show that the stochastic Ito integral $ \textrm{Ito}\int_0^tdt' \cdots $ in Eq.~\eqref{eq:PW_expansion_Ito} is equivalent to the stochastic Stratonovich integral $ \int_0^tdt' \cdots $  obeying conventional integration rules~\cite{Asheichyk2020, Cugliandolo2017, Gardiner2010, Wynants2010}. We can hence write
\begin{equation}
\mathcal{W}^{(\dot{\gamma})} - \mathcal{W} = \mathcal{W}\frac{\dot{\gamma}}{2k_{\textrm{B}}T}\int_0^tdt'\sum_{i=1}^N\mathcal{X}_{ix}(t')y_i(t') + \mathcal{O}(\dot{\gamma}^2).
\label{eq:PW_expansion_Str}
\end{equation}
From Eq.~\eqref{eq:PW_expansion_Str}, the linear response  relation follows immediately:
\begin{equation}
\Delta A^{(\dot{\gamma})} = \frac{\dot{\gamma}}{2k_{\textrm{B}}T}\int_0^tdt'\left\langle A(t)\sum_{i=1}^N\mathcal{X}_{ix}(t')y_i(t')\right\rangle.
\label{eq:GKU_der1}
\end{equation}

Next, we use the time reversal symmetry of the equilibrium linear response: the term containing $ \sum_{i=1}^Nv_{ix}y_i $ in Eq.~\eqref{eq:GKU_der1} is antisymmetric with respect to the time reversal, and it equals the other term which is time symmetric~\cite{Baiesi2009_1, Baiesi2009_2, Baiesi2010, Basu2015}. Therefore, the response can be given by two times either of the terms. For the Green-Kubo relation, we need the time-symmetric one:
\begin{align}
\notag &\Delta A^{(\dot{\gamma})} = \frac{\dot{\gamma}}{k_{\textrm{B}}T}\int_0^tdt'\\
&\times \left\langle A(t)\sum_{i=1}^N\left[m\dot{v}_{ix}(t')- F^{\textrm{int}}_{ix}(t') - F^{\textrm{ext}}_{ix}(t')\right]y_i(t')\right\rangle.
\label{eq:GKU_der2}
\end{align} 

Finally, we use partial integration for the first term in Eq.~\eqref{eq:GKU_der2}:
\begin{align}
\notag & \int_0^tdt'\sum_{i=1}^N\dot{v}_{ix}(t')y_i(t') = -\int_0^tdt'\sum_{i=1}^Nv_{ix}(t')v_{iy}(t')\\
& +\sum_{i=1}^Ny_i(t)v_{ix}(t) - \sum_{i=1}^Ny_i(0)v_{ix}(0).
\label{eq:GKU_der3}
\end{align} 
Substituting Eq.~\eqref{eq:GKU_der3} into Eq.~\eqref{eq:GKU_der2} and identifying the stress tensor~\eqref{eq:sigma}, we obtain formula~\eqref{eq:LR_GKU}.

\onecolumngrid
\section{The linear response of a single underdamped particle confined in a harmonic potential}
\label{app:resp1}
The response of a single underdamped particle can be evaluated analytically~\cite{Rzehak2003, Holzer2010, Lander2012, Kahlert2012}. This appendix complements the results of 
Refs.~\cite{Rzehak2003, Holzer2010, Lander2012, Kahlert2012} with the transient response. A derivation of the expressions given here as well as a check of formulas~\eqref{eq:LR_GKU} 
and~\eqref{eq:LR_RK} can be found in Ref.~\cite{Asheichyk2020}.

\begin{figure*}[!t]
\begin{tabular}{cc}
\includegraphics[width=0.49\linewidth]{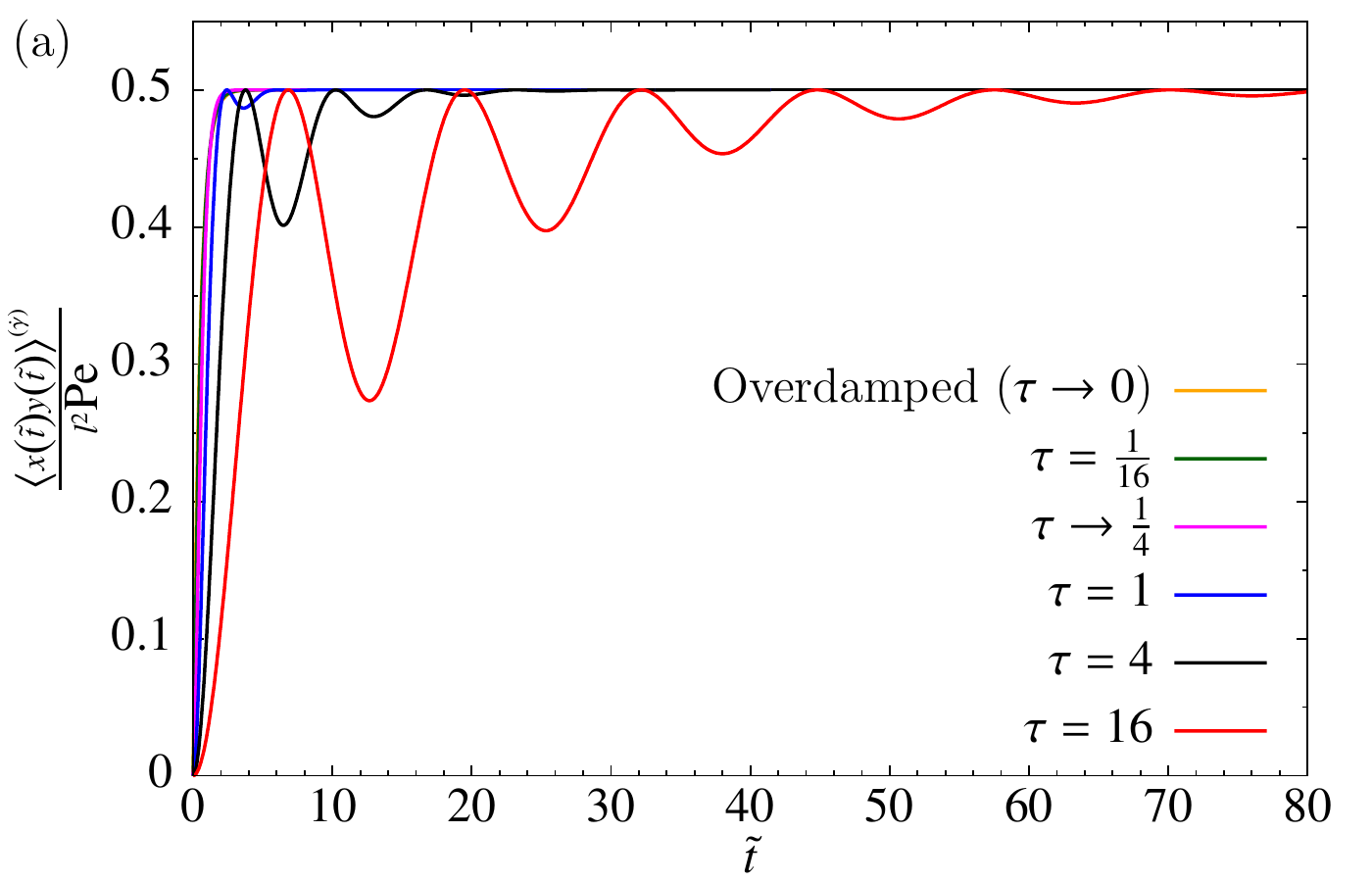}
&
\includegraphics[width=0.49\linewidth]{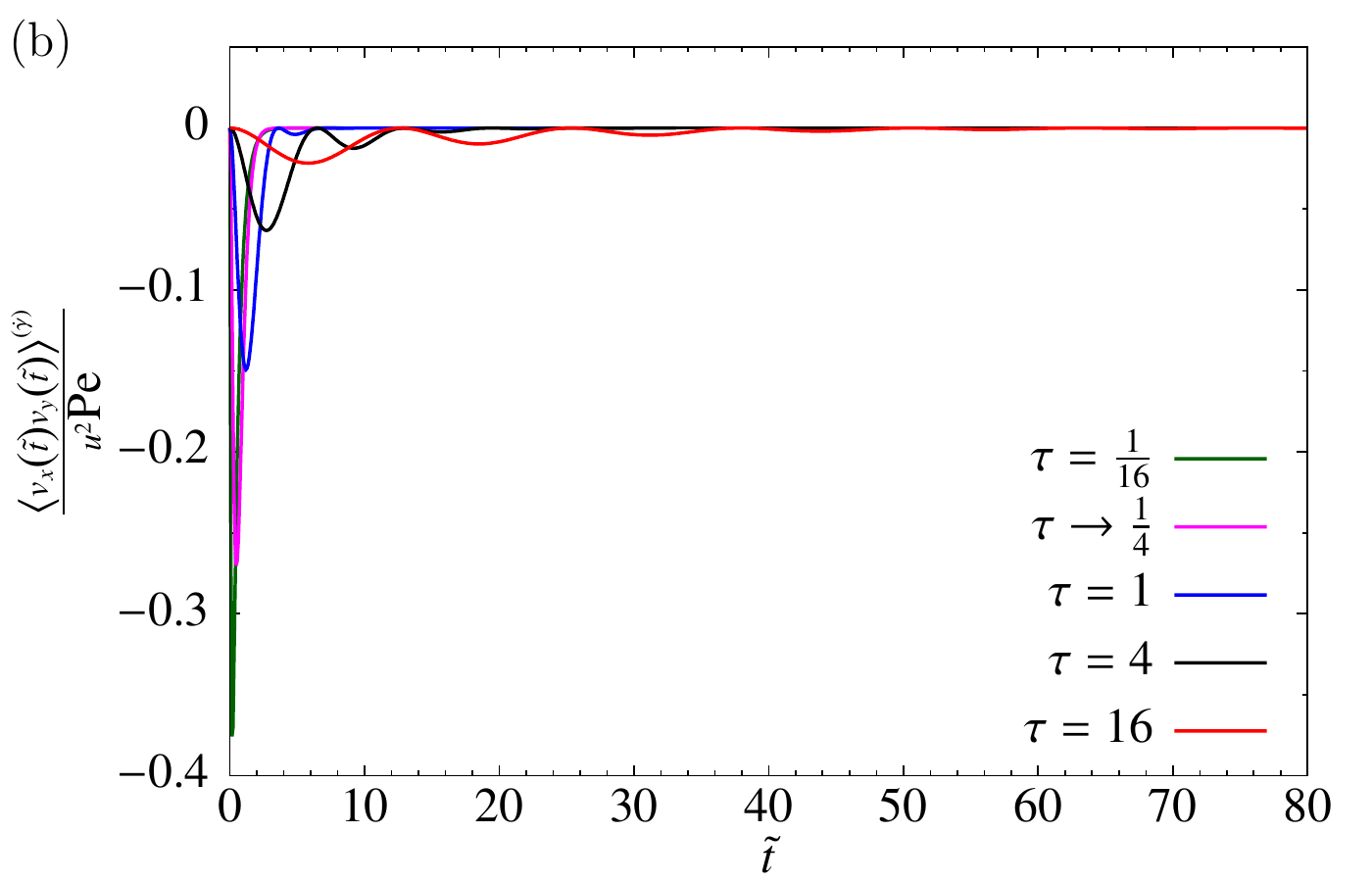}\\
\includegraphics[width=0.49\linewidth]{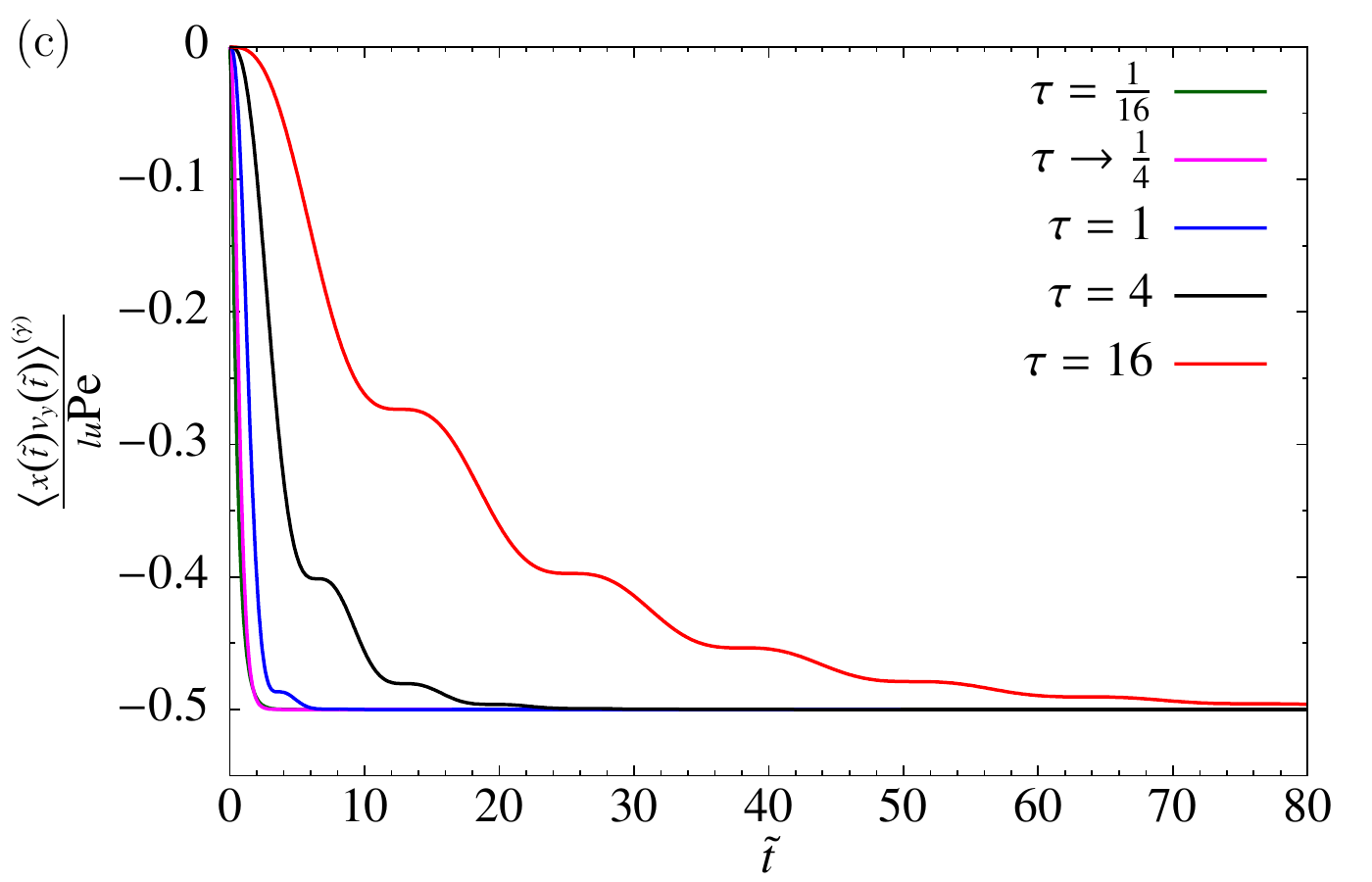}
&
\includegraphics[width=0.49\linewidth]{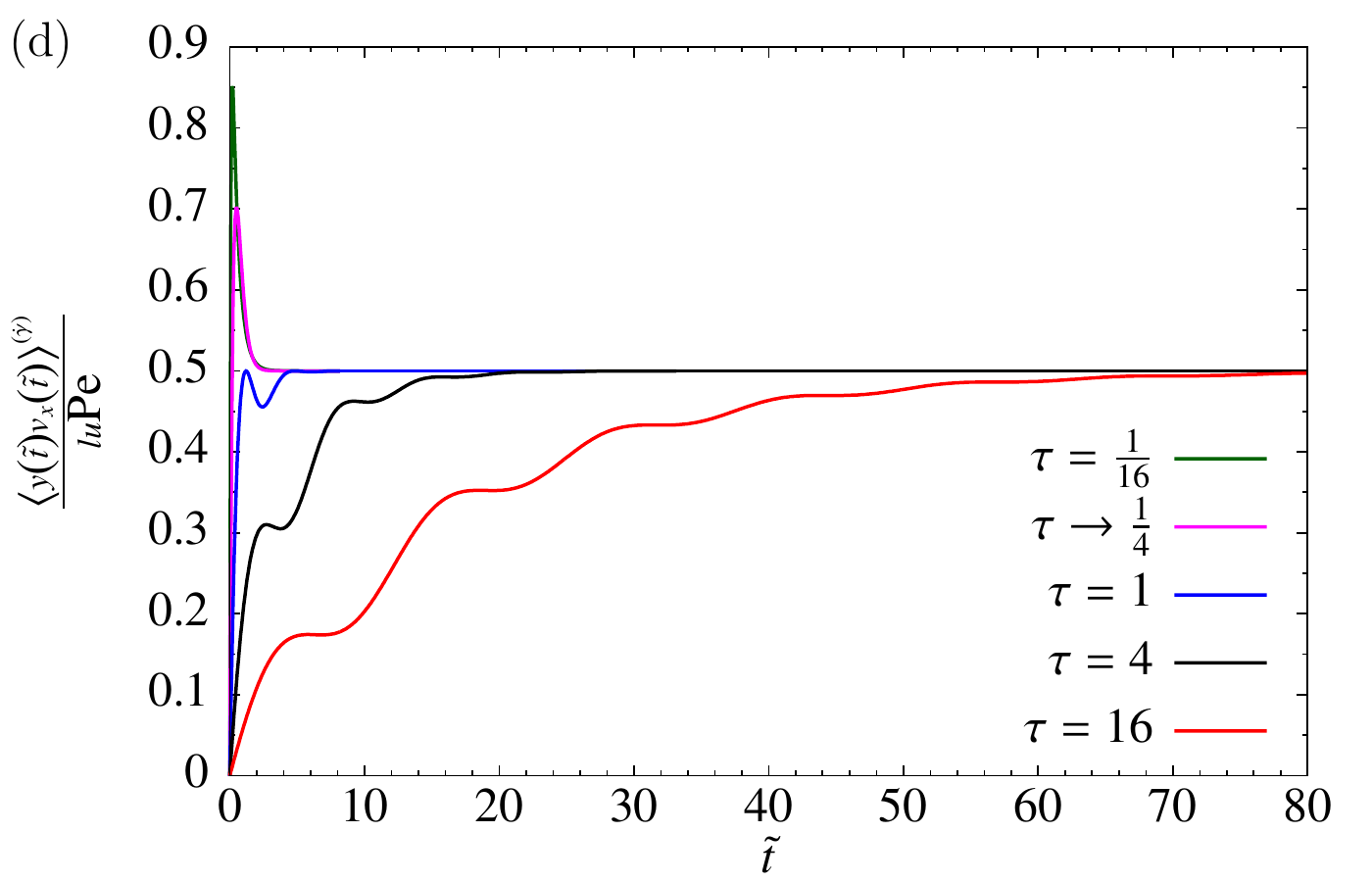}
\end{tabular}
\caption{\label{fig:AE}Rescaled response to shear flow for $ A = xy $ [(a), Eq.~\eqref{eq:AE_xy_direct}], $ A = v_xv_y $ [(b), Eq.~\eqref{eq:AE_vxvy_direct}], $ A = xv_y $ 
[(c), Eq.~\eqref{eq:AE_xvy_direct}], and $ A = yv_x $ [(d), Eq.~\eqref{eq:AE_yvx_direct}] of a single underdamped particle confined in a harmonic trap as a function of 
rescaled time $ \tilde{t} \equiv \frac{t}{\tau_k} = \mu kt $ after the flow is applied. The results are shown for different values of a characteristic ratio 
$ \tau \equiv \frac{\tau_m}{\tau_k} $, which is the ratio between the inertia time $ \tau_m = \mu m $ and the trap relaxation time $ \tau_k = \frac{1}{\mu k} $. Different values of $ \tau $ 
correspond to different values of $ m $, while $ \mu $ and $ k $ are assumed to be fixed and finite; $ \tau \to 0 $ is the overdamped limit. 
$ \textrm{Pe} \equiv \dot{\gamma}\tau_k = \frac{\dot{\gamma}}{\mu k} $, $ l^2 \equiv k_{\textrm{B}}T\mu\tau_k = \frac{ k_{\textrm{B}}T}{k} $, and 
$ u^2 \equiv \frac{l^2}{\tau_k^2} = k_{\textrm{B}}T\mu^2k $ are the Peclet number, the unit of squared length, and the unit of squared velocity, respectively.}
\end{figure*}

The responses for $ A = xy $, $ A = v_xv_y $, $ A = xv_y $, and $ A = yv_x $ read (we note that $ \left\langle A \right\rangle = 0 $ for all considered $ A $, i.e., $ \Delta A^{(\dot{\gamma})} = {\left\langle A(t) \right\rangle}^{(\dot{\gamma})}$)
\begin{align}
\notag {\left\langle x(t)y(t) \right\rangle}^{(\dot{\gamma})} = & \ \frac{2\dot{\gamma}k_{\textrm{B}}T\mu}{1-4\mu^2km}\Bigg\{\left(\frac{\mu m}{1-\sqrt{1-4\mu^2km}}\right)^2\left(1-e^{-\frac{1}{\mu m}\left(1-\sqrt{1-4\mu^2km}\right)t}\right)\\
& +\left(\frac{\mu m}{1+\sqrt{1-4\mu^2km}}\right)^2\left(1-e^{-\frac{1}{\mu m}\left(1+\sqrt{1-4\mu^2km}\right)t}\right) - \frac{m}{2k}\left(1-e^{-\frac{t}{\mu m}}\right)\Bigg\},
\label{eq:AE_xy_direct}
\end{align}
\begin{equation}
{\left\langle v_x(t)v_y(t) \right\rangle}^{(\dot{\gamma})} = \frac{\dot{\gamma}k_{\textrm{B}}T\mu}{2\left(1-4\mu^2km\right)}\Bigg\{\left(1-e^{-\frac{1}{\mu m}\left(1-\sqrt{1-4\mu^2km}\right)t}\right) +\left(1-e^{-\frac{1}{\mu m}\left(1+\sqrt{1-4\mu^2km}\right)t}\right) - 2\left(1-e^{-\frac{t}{\mu m}}\right)\Bigg\},
\label{eq:AE_vxvy_direct}
\end{equation}
\begin{align}
\notag {\left\langle x(t)v_y(t) \right\rangle}^{(\dot{\gamma})} = & \ -\frac{\dot{\gamma}k_{\textrm{B}}T\mu}{1-4\mu^2km}\Bigg\{\frac{\mu m}{1-\sqrt{1-4\mu^2km}}\left(1-e^{-\frac{1}{\mu m}\left(1-\sqrt{1-4\mu^2km}\right)t}\right)\\
& +\frac{\mu m}{1+\sqrt{1-4\mu^2km}}\left(1-e^{-\frac{1}{\mu m}\left(1+\sqrt{1-4\mu^2km}\right)t}\right) - 2\mu m\left(1-e^{-\frac{t}{\mu m}}\right)\Bigg\},
\label{eq:AE_xvy_direct}
\end{align}
\begin{align}
\notag {\left\langle y(t)v_x(t) \right\rangle}^{(\dot{\gamma})} = & \ -\frac{\dot{\gamma}k_{\textrm{B}}T\mu}{1-4\mu^2km}\Bigg\{\frac{\mu m}{1-\sqrt{1-4\mu^2km}}\left(1-e^{-\frac{1}{\mu m}\left(1-\sqrt{1-4\mu^2km}\right)t}\right)\\
& +\frac{\mu m}{1+\sqrt{1-4\mu^2km}}\left(1-e^{-\frac{1}{\mu m}\left(1+\sqrt{1-4\mu^2km}\right)t}\right) - \frac{1-2\mu^2km}{\mu k}\left(1-e^{-\frac{t}{\mu m}}\right)\Bigg\}.
\label{eq:AE_yvx_direct}
\end{align}
These results are valid for any $ \dot{\gamma} $, i.e., the linear response for a single particle and the considered observables equals the total response, in contrast to a many-particle 
system (see Fig.~\ref{fig:NL_GD_lin} and Ref.~\cite{Asheichyk2019_2}). In order to visualize Eqs.~\eqref{eq:AE_xy_direct},~\eqref{eq:AE_vxvy_direct},~\eqref{eq:AE_xvy_direct}, and~\eqref{eq:AE_yvx_direct}, we make the 
equations dimensionless~\cite{Asheichyk2019_2} and plot them in Fig.~\ref{fig:AE}; definitions of the parameters are given in the caption.

\twocolumngrid
\section{Standard deviation for a single overdamped particle}
\label{app:var}
This appendix gives analytical results for the standard deviation (SD) of the response relations in case of a single overdamped particle. We consider the same system as in Sec.~\ref{sec:Investigation}, and choose $ A = xy $. As it is shown in Subsec.~\ref{subsec:Variance}, SD is different for different response relations. It is defined as
\begin{equation}
\sigma = \sqrt{\left\langle B^2\right\rangle - {\left\langle B\right\rangle}^2},
\label{eq:var_def}
\end{equation}
where, for the chosen system and observable,
\begin{subequations}
\begin{alignat}{4}
& B_{\textrm{Work}} = \frac{\dot{\gamma}}{k_{\textrm{B}}T\mu}\int_0^tdt'x(t)y(t)y(t')\dot{x}(t'),\label{eq:B_TA}\\
& B_{\textrm{RF}} = \frac{\dot{\gamma}}{2k_{\textrm{B}}T}\int_0^tdt'x(t)y(t)y(t')f_x(t'),\label{eq:B_RF}\\
& B_{\textrm{GK}} = \frac{\dot{\gamma}k}{k_{\textrm{B}}T} \int_0^tdt'x(t')y(t')x(0)y(0),\label{eq:B_GK}\\
& B_{\textrm{sFDT}} = \frac{\dot{\gamma}}{2k_{\textrm{B}}T\mu}\left[x^2y^2 - x(t)y(t)x(0)y(0)\right],\label{eq:B_RK}
\end{alignat}
\end{subequations}
for formulas~\eqref{eq:LR_TA},~\eqref{eq:LR_RF},~\eqref{eq:LR_GK}, and~\eqref{eq:LR_RK}, respectively. The difference in SDs originates from the term $ \left\langle B^2\right\rangle $ in Eq.~\eqref{eq:var_def}, while the result for $ {\left\langle B\right\rangle}^2 $ is the same for all the aforementioned $ B $ and equals~\cite{Asheichyk2019_1, Asheichyk2020}
\begin{equation}
{\left\langle B\right\rangle}^2 = \frac{\dot{\gamma}^2(k_{\textrm{B}}T)^2}{4\mu^2k^4}\left(1-e^{-2\mu kt}\right)^2.
\label{eq:Bav2}
\end{equation}

In order to evaluate $ \sigma $ for $ B $ in Eq.~\eqref{eq:B_TA}, we use the Langevin equation to replace $ \dot{x} $ with $ \dot{x} = -\mu kx + \mu f_x $.  The double integrals containing $ f_x $ are evaluated using standard (Stratonovich) integration rules. The corresponding SDs computed in numerical simulations can hence differ from the analytical results provided below [Eqs.~\eqref{eq:varTA} and~\eqref{eq:varRF}] if the discretization scheme different from the Stratonovich one is used; however, the scaling with $ \sqrt{t} $ for large $ t $ should not be affected by this issue.

Here are our results:
\begin{subequations}
\begin{alignat}{4}
&\sigma_{\textrm{Work}} = \frac{\dot{\gamma}k_{\textrm{B}}T}{2\mu k^2}\Big(4\mu kt + 13 -16e^{-2\mu kt} + 3e^{-4 \mu kt}\Big)^{\frac{1}{2}},\label{eq:varTA}\\
\notag & \sigma_{\textrm{RF}} = \frac{\dot{\gamma}k_{\textrm{B}}T}{2\mu k^2}\Big(2\mu kt + 9 - 8\mu kte^{-2\mu kt} - 12e^{-2\mu kt}\\
& \ \ \ \ \ \ \ \ \ \ \ \ \ \ \ \ \ \ \ + 3e^{-4 \mu kt}\Big)^{\frac{1}{2}},\label{eq:varRF}\\
\notag & \sigma_{\textrm{GK}} = \frac{\dot{\gamma}k_{\textrm{B}}T}{2\mu k^2}\Big(4\mu kt + 9 - 16\mu kte^{-2\mu kt} - 12e^{-2\mu kt}\\
& \ \ \ \ \ \ \ \ \ \ \ \ \ \ \ \ \ \ \ + 3e^{-4 \mu kt}\Big)^{\frac{1}{2}},\label{eq:varGK}\\
& \sigma_{\textrm{sFDT}} = \frac{\dot{\gamma}k_{\textrm{B}}T}{2\mu k^2}\Big(9 -12e^{-2\mu kt} + 3e^{-4 \mu kt}\Big)^{\frac{1}{2}},\label{eq:varRK}
\end{alignat}
\end{subequations}
corresponding to response relations~\eqref{eq:LR_TA},~\eqref{eq:LR_RF},~\eqref{eq:LR_GK}, and~\eqref{eq:LR_RK}, respectively. In the limit of large time, $ \mu kt \gg 1 $, SD scales as $ \sqrt{t} $ for the work formula, random force formula, and Green-Kubo relation, whereas it approaches a constant for sFDT. Considering that the steady state is reached at $ \mu kt \approx 5 $ [see Eq.~\eqref{eq:Bav2}], we conclude from Eqs.~\eqref{eq:varTA},~\eqref{eq:varRF},~\eqref{eq:varGK}, and~\eqref{eq:varRK} that, in the steady state, $ \sigma_{\textrm{sFDT}} < \sigma_{\textrm{RF}} < \sigma_{\textrm{GK}} < \sigma_{\textrm{Work}} $. Note that this inequality differs from what we found numerically (using the Euler discretization) in Subsec.~\ref{subsec:Variance} for a many-particle underdamped system, because of the aforementioned dependence of $ \sigma_{\textrm{Work}} $ and $ \sigma_{\textrm{RF}} $ on the discretization procedure.

\section{Response to the shear potential}
\label{app:respUptb}
When the system of particles confined in a harmonic trap $ U^{\textrm{ext}} = \frac{k}{2}\sum_{i=1}^N(x_i^2+y_i^2) $ is perturbed by the shear potential~\eqref{eq:Ushear}, the resulting external potential becomes
\begin{equation}
U^{\textrm{ext}} = \frac{k}{2}\sum_{i=1}^N\left(x_i^2+y_i^2\right) - \frac{\dot{\gamma}}{2\mu}\sum_{i=1}^Nx_iy_i.
\label{eq:U}
\end{equation}
Each particle hence feels the external potential
\begin{equation}
u = \frac{k}{4}\left[(1-\delta)(x+y)^2 + (1+\delta)(x-y)^2\right],
\label{eq:u}
\end{equation}
with $ \delta = \frac{\dot{\gamma}}{2\mu k} $.

\begin{figure}[!b]
\includegraphics[width=1.0\linewidth]{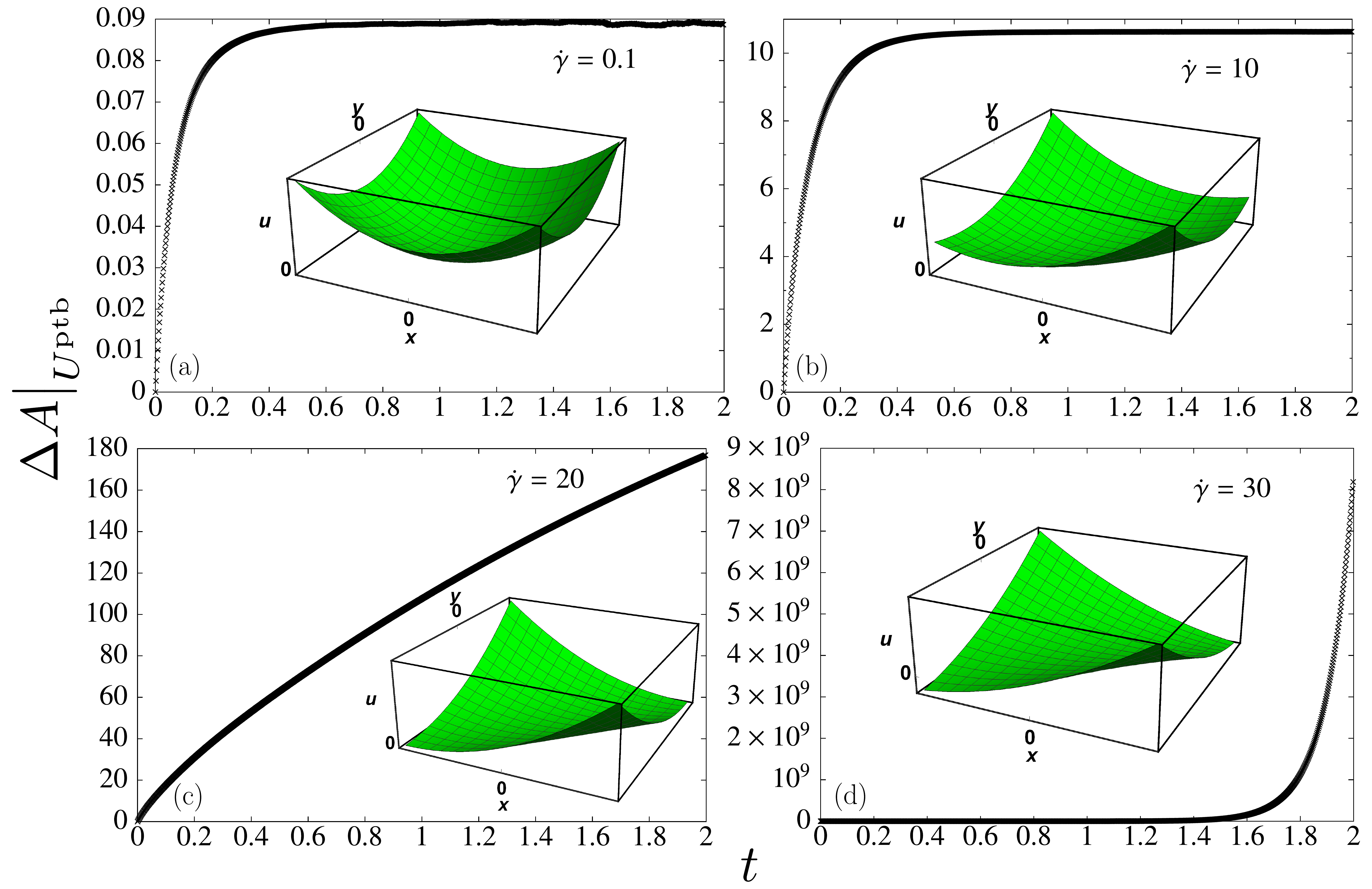}
\caption{\label{fig:respUptb}Response to the shear potential~\eqref{eq:Ushear} as a function of time for different shear rates. The insets illustrate the sum of external potential and perturbation potential. The system, observable, and parameters are the same as in Fig.~\ref{fig:NL_GD_lin}.}
\end{figure}

If $ \delta < 1 $ (i.e., $ \dot{\gamma} < 2\mu k $), the point $ (0, 0) $ is a minimum of $ u $ [see the insets of Figs.~\ref{fig:respUptb}(a) and~\ref{fig:respUptb}(b)], i.e., there is a well-defined equilibrium. Therefore, $ \Delta A\big|_{U^{\textrm{ptb}}} $ converges to a steady-state value [see Figs.~\ref{fig:respUptb}(a) and~\ref{fig:respUptb}(b)]. 

If $ \delta = 1 $ (i.e., $ \dot{\gamma} = 2\mu k $), $ x = y $ is a \enquote{minimum line} [see the inset of Fig.~\ref{fig:respUptb}(c)], i.e., the particles tend to go along the diagonal $ x = y $ with no confinement on that way. Therefore, $ \Delta A\big|_{U^{\textrm{ptb}}} $ diverges [see Fig.~\ref{fig:respUptb}(c)]. 

Finally, if $ \delta > 1 $ (i.e., $ \dot{\gamma} > 2\mu k $), $ (0, 0) $ is a saddle point, with the line $ x = y $ being unstable yet preferable to go along [see the inset of Fig.~\ref{fig:respUptb}(d)]. Therefore, $ \Delta A\big|_{U^{\textrm{ptb}}} $ diverges exponentially  [see Fig.~\ref{fig:respUptb}(d)].

\end{appendix}


% FOOTER_end ---------------------------------------------

\end{document}